\documentclass{scrartcl}
\usepackage{url}
\usepackage{graphicx}

\usepackage{amssymb}

\begin{document}

\title{BeSpaceD}
\subtitle{Towards a Tool Framework and Methodology for the Specification and Verification of Spatial Behavior of Distributed Software Component Systems}

\author{
Jan Olaf Blech and Heinz Schmidt
}
\date{RMIT University, Melbourne, Australia}
\maketitle
\begin{abstract}
In this report, we present work towards a framework for modeling and checking behavior of spatially
distributed component systems. Design goals of our framework are the
ability to model spatial behavior in a component oriented, simple and
intuitive way, the possibility to automatically analyse and verify
systems and integration possibilities with other modeling and
verification tools. 
We present examples and the verification
steps necessary to prove properties such as range coverage or the absence of collisions between
components and technical details.
\end{abstract}

\section{Introduction}

Ensuring that distributed embedded systems meet their requirements is an important
and well-studied topic in industry and research. With growing system
sizes and the combination of independently developed systems into
larger systems of systems the
interplay of different subsystems becomes an important aspect of
study. When regarding embedded controllers integrated into other
devices such as ro\-bots or cars, the software running on these systems does have an
impact on the physical environment.  Potential hazards resulting from
incorrect software can lead to damaged property, injuries or even loss of human lives. When
combining systems that have a physical impact on an environment, one
may have to study the subsystems together, since one can detect some potential hazards,
e.g., two cars driving in opposite directions, but on a single lane
only in the combined system.

Different techniques have been introduced for modeling and checking of
cyber-physical systems and their properties. These comprise differential equations, automata
and different notions of time such as continuos or discrete time and others. Based
on these modeling approaches, tools for checking properties have been
developed.
The software engineering community has been
studying the design and architecture of component-based systems for
decades. Specification formalisms like automata and message sequence
charts are frequently used to describe expected behavior of
systems. Different means for checking these specifications are available.

In this report, we present work towards a unified approach to
cyber-physical systems and component-based software development. We
motivate a new framework 
for spatial cyber-physical behavior, communication and interaction
between different components. 
Our framework is especially suited for large-scale widely distributed systems with limited interactions between components but also allows the modeling of the precise geometric behavior of components in space. Several components may be distributed over a large space
  and have distinct features of communication, spatial and internal
  behavioral aspects. We describe a process
  for checking properties of our models. In particular, we are
  focusing on spatial aspects: a component features different
  dimensions of possible non-deterministically occupied space, e.g., characterizing its physical dimensions,
  and the reach of sensors and communication devices.
We study modeling and verification scenarios which are
characterized by a discrete notion of time that features 
  {\it time points} and {\it time intervals} between them. Time points are partially ordered. This allows for the
  modelling of synchronous and asynchronous systems with distinct
  synchronisation points between different components.

Some preliminary ideas to this report are presented in \cite{issec}. An
application of our framework in an IDE for the development of reactive
probabilistic systems is featured among other content in \cite{fesca2014}.

\paragraph{Overview}

Section~\ref{sec:relwork} discusses related work. Modeling of systems
with our framework and examples are presented in
Section~\ref{sec:model}. Section~\ref{sec:reason} describes ways to
reason about our models and perform
verification. Section~\ref{sec:impl} discusses our implementation in
Scala. Conclusion and
future work is presented in Section~\ref{sec:concl}.

\section{Related Work}
\label{sec:relwork}

Work that is relevant to this paper has been done in areas such as formal
logic and process algebras, hybrid-systems, robotics and formal
methods for component-based software engineering.

\paragraph{Formal Logic and Process Algebraic Approaches}
The handbook of spatial logic \cite{hosl} discusses spatial logics,
related algebras and applications. The book covers a large spectrum
not only limited to computer science.
A process algebra like formalism for describing and reasoning about
spatial behavior has been introduced in \cite{cardelli03} and
\cite{cardelli04}. Process algebras come with a precise formal
semantics definition and are aimed towards the specification of highly
parallel systems. Here, disjoint logical spaces are
represented in terms of expressions by bracketing structures and carry
or exchange  concurrent  process representations. A model checking
tool for process algebra like spatial behavioral specifications is
presented in \cite{slmc}. A graph-based technique for the verification
of spatial properties of finite $\pi$-calculus fragments is introduced
in \cite{gadducci}.

For results on spatial interpretations see, e.g.,
\cite{hirschkoff}. Many aspects of spatial logic are in general
undecidable. A quantifier-free rational fragment of ambient logic
(corresponding to regular language constraints), however, has been shown to be decidable in \cite{zilio}.

Special modal logics for spatio-temporal reasoning go back to the
seventies. The Region Connection Calculus (RCC) \cite{Bennett}  includes
spatial predicates of separation. For example RCC features predicates
indicating that regions do not share points at all, points on the
boundary of regions are shared, internal contact  where one region is included
and touches on the boundary of another from the inside, proper overlap of
regions,
and proper inclusion.
In addition \cite{Bennett}  features an overview of the relation of
these logics to various Kripke-style modal logics, reductions of
RCC-style fragments to a minimal number of topological predicates,
their relationship to interval-temporal logics and decidability.

\paragraph{Hybrid-systems}
The area of hybrid systems has seen the development of different
tools for reasoning and verification. SpaceEx \cite{frehse2011} allows
the modeling of continuos hybrid systems based on hybrid automata. It
can be used for computing overapproximations of the space occupied by
an object moving in time and space. 
Additionally, it is possible to model spatial behavior in more general
purpose oriented verification tools in Hybrid systems (e.g., \cite{keymaera}).

\paragraph{Robotics}
Related to our work is the work on path planning for robots (e.g.,
\cite{robots1,robots2}). In our work, however, we are concentrating on
checking existing properties of systems rather than optimization or
discovery of new possible paths. Collision detection for robots in
combination with motion planning has been studied for a long time,
see, e.g., \cite{collision} and \cite{canny}. Strongly related to
motion planning is the task of efficiently handling geometric reasoning. On this geometric interpretation level, techniques have been
investigated to structure the tasks of detecting possible
inference between geometric objects (e.g., \cite{obb-tree} and \cite{geoinf2}) for
efficient analysis. 

\paragraph{Data Models and GIS like Services}
Data models for cyberphysical infrastructure in construction, plant
automation and transport -- domains that we are
aiming for in this paper -- have been studied in the past. Unlike in
this paper, many of the existing real-world applications are aligned
towards a geometric representation of components and are typically based on so-called 2.5 dimensional
GIS (Geographic Information System) representations where the 3rd dimension $z=f(x,y)$ is represented
as a function $f$ of the 2 dimension $x$ and $y$ coordinates. This
\cite{apel} limits the geometric, topological and information
retrieval use of such models. True three dimensional modeling is far from
common practice \cite{friedman}. 
Our approach is not limited to a
particular geometric representation, coordinate or dimension system.
Future extensions of interest include consideration of standards such as the Web 3D Services and the Sensor Web Enablement Architecture of the Open Geospatial Consortium~\footnote{\url{http://www.opengeospatial.org}}, visualisation and decision support \cite{weaver} and efficient data structures for fast meta reasoning and presenting subproblems of choice to specialist solvers used in our examples.

\paragraph{Component-based Systems}
Multiple formalisms to describe
component behavior can be used as specification basis for our methodology. For example the UML 2.0 \cite{uml} standard provides
message sequence charts and state machines that can be used to
describe component behavior and interactions. These can be used to
derive information about shared lifetimes of components and behavior.

The notion of time and component interaction used in this paper is compatible with the Petri-nets induced notion. This is used, e.g., in the BIP framework
\cite{bip1} for modeling of distributed asynchronous
systems. Invariants are used as an intermediate step for verification
in the BIP context \cite{dfinder2}. Invariants are also used as an
intermediate representation of components in this work.

We have been investigating mathematical models of behavior in space in
previous work
\cite{shape}, which also features a survey on work related to these
models. A prestudy of the work described in this paper is presented in \cite{issec}. Another software
architectural model was proposed in \cite{radl}. Specification and reasoning on existing software
component systems \cite{blech1,blech2}  in the context of a type
system for behavior has also been examined by
us and is an important direction of future extensions of this
paper. In this work we aim towards a unified view of component system
aspects and verification and
the introduction of space as an additional feature. 
 Symbolically reasoning about invariants of asynchronous distributed systems, which is a part of our verification methodology in this paper, has been studied by us in \cite{bipcoq}.

\section{Modeling of Spatial Behavior}
\label{sec:model}
This section presents the ingredients of our modeling framework for spatial
behavior and examples.

\subsection{Guiding Examples}
Here, we present four guiding examples, concurrent window cleaning, 
moving forklifts, communicating cars on a road network, and rotating robots.

\paragraph{Concurrent Window Cleaning}
\begin{figure}
\centering 
\includegraphics[width=0.3\textwidth]{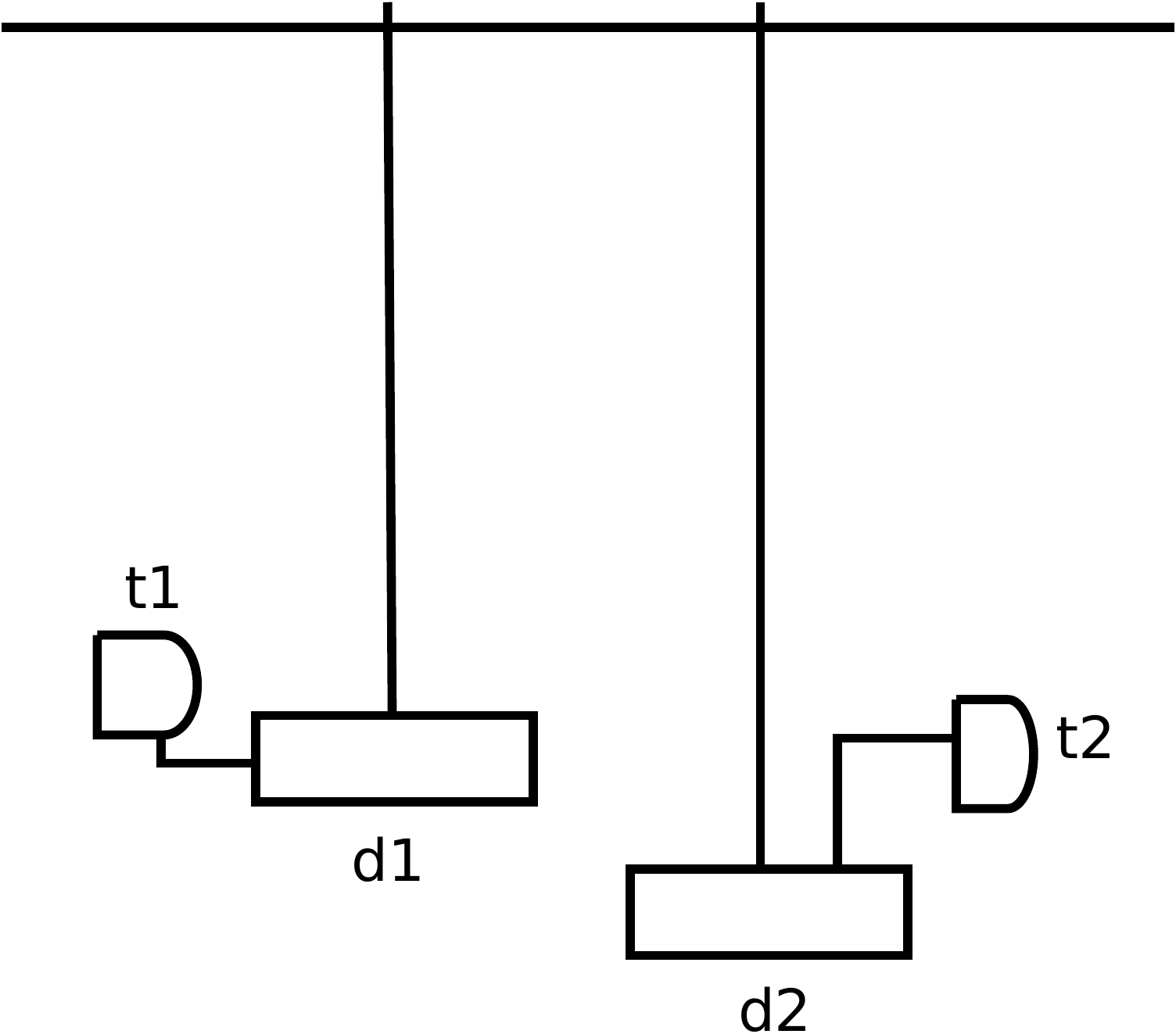}
\caption{Concurrent window cleaning}
\label{fig:ex1}
\end{figure}
Figure~\ref{fig:ex1} shows a concurrent window cleaning
system. Platforms $d_1$, $d_2$ are attached to a mobile device on a
roof. They can be moved horizontally by moving the mobile device on
the roof and vertically through a rope. 
Attached to each platform is a robot arm: $t_1$ and $t_2$. It can have
different positions relative to its platform, but only within a limited
range.
Furthermore, we have some kind of internal state 
for each platform.

The behavior of each robot is controlled by a program. This program
implies behavior which is potentially non-deterministic and may depend
on other robots or external events. The behavior is characterized by
spatial aspects, the movement of the robot, communication aspects
(e.g., interactions with other robots or some external controlling
device), and internal state changes.

We are interested in simulating possible window cleaning
scenarios. Robots have local states and when they interact with
another robot their actions may undergo a synchronization. This
synchronization does not need to be global, i.e., in case of many
robots cleaning a window, the synchronisation does not have to be shared with all
robots. For this reason, we only have a partial order of time. Each
element of this partial order is called a time point.

One aspect that we are interested in, is whether a certain system state implies a collision. 
To do this, we could examine the position of each platform and each robot arm. Based on this, we could calculate the exact space boundaries that each device uses. 
An alternative way is to use an abstraction and use an
overapproximation of the space used. More coarse grained abstractions may only have to take the position of the platforms into account. 

\begin{figure}
\centering
\includegraphics[width=0.45\textwidth]{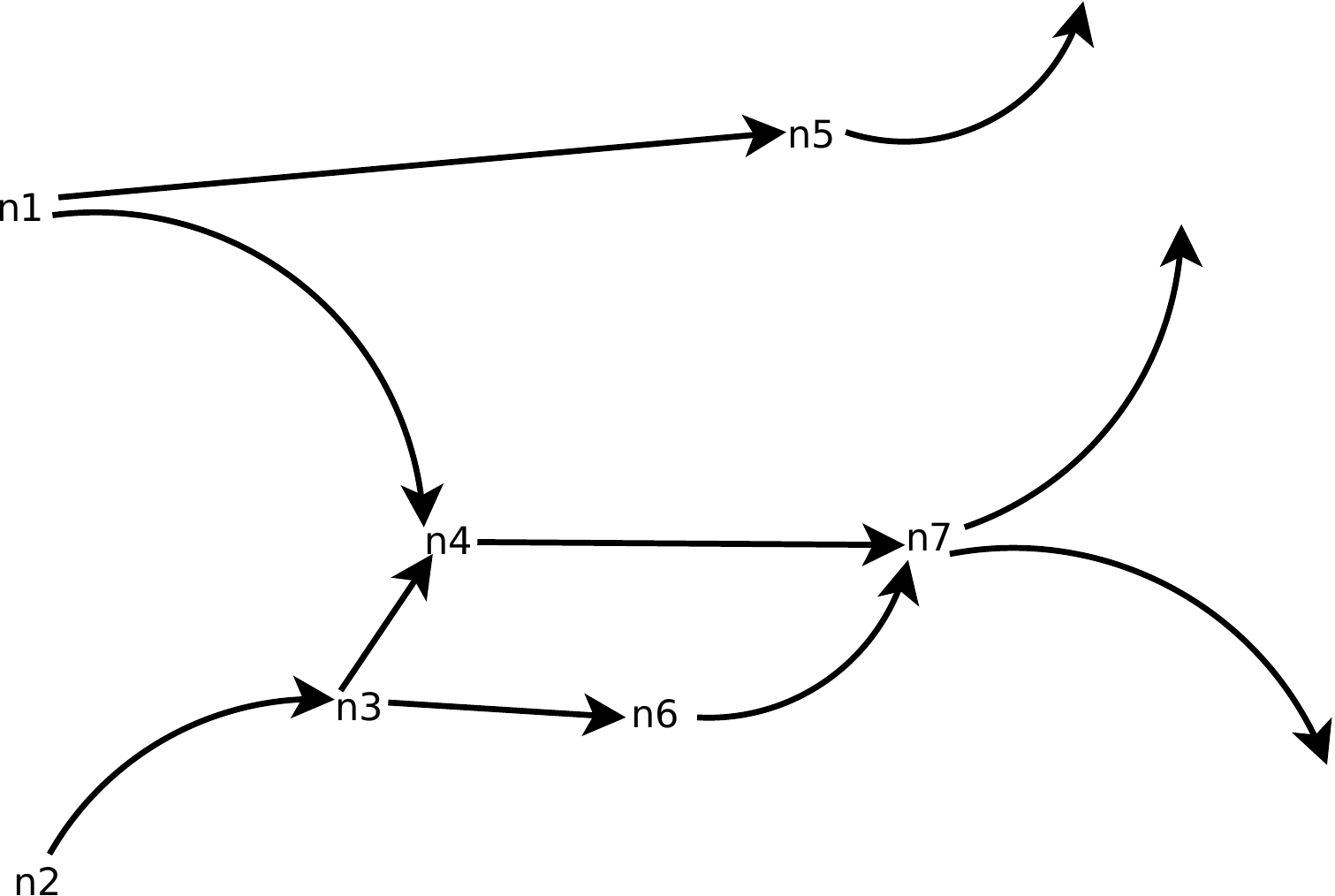}
\caption{Forklifts on a network}
\label{fig:newex}
\end{figure}

\paragraph{Forklifts}
Figure~\ref{fig:newex} shows a second  example. Suppose that two forklifts want
to drive on the pictured network of paths and perform tasks.
The behavior of the two forklifts is written as follows:
\begin{itemize}
\item Forklift 1 starts at n1. It can continue either to n5 or via n4 to
  n7 in order to exit the system. This is unknown at the time the system starts.
\item  Forklift 2 starts at n2 and continues to n3. If it can detect
  that n4 is occupied at will continue to n6 and than n7. If n4 is
  unoccupied it will continue to n7 via n4. Forklift 2 will exit the
  system via n7.
\end{itemize}
Further constraints apply:
\begin{itemize}
\item
In addition to the movement in the paths network, each forklift
occupies a certain amount of space and drives at a certain speed. For
distinct time points it is possible to determine an area in which the
forklift will reside.
\item
Additionally each forklift has a locally limited ability to detect
obstacles and other forklifts.
\end{itemize}

Possible questions apply:
1) Can collisions occur?
2) Are forklifts delayed by other forklifts that can be detected and avoided?

\begin{figure}[htb]
\centering
\includegraphics[width=0.47\textwidth]{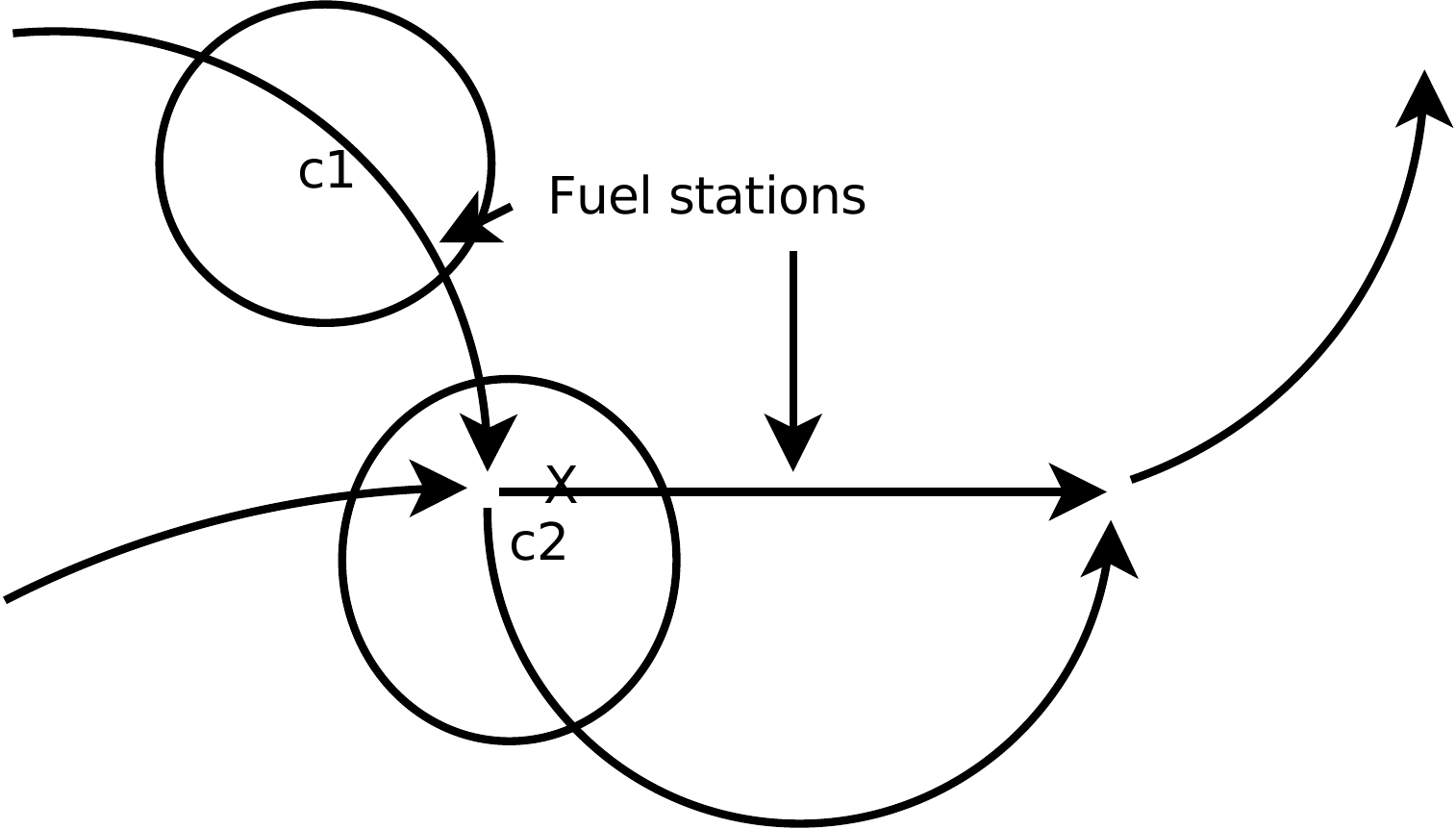}
\caption{Communicating cars}
\label{fig:ex2}
\end{figure}

\paragraph{Driving Cars}
Figure~\ref{fig:ex2} shows two cars $c1$ and $c2$ driving on a network of roads. The
road network is formalized as a graph. Cars have a limited ability of
locally communicating with each other, iff they are within a certain
distance of each other. This distance is indicated by a circle around
the car in the figure.
A car has a local state. This is defined by the following ingredients:
\begin{itemize}
\item
The road section it is currently driving on: the edge of the corresponding graph.
\item
The position on the road, e.g., indicating the progress on the road
section and the direction it is facing.
\item Aspects of the internal state that encapsulates the parts of a state that
  are truly local to the car like the amount of fuel left. 
\item Communication aspects of an internal state, e.g., states that are reached throughout
  an ongoing communication effort with other cars according to a protocol.
\end{itemize}
One can see, that spatial, truly internal and communication aspects
are encoded in the states.

For calculating the distance between two cars travelling on different
roads, we need an interpretation. We take the road edges and the
position on the road into account. 
The topological graph information needs to be interpreted in a
geometrical way. For example, we can retrieve data from a geographic
information system to get the exact location on earth that allows us
to compute the distance between the two cars. Alternatively a car may
itself keep track of the actual coordinates within its
state. Furthermore, we need a shared time points between the two cars
so that we have access to their positions at the same time.

The local communication possibilities of the cars may be used for
various purposes. For example, a car might inform another car that
a road has been blocked (indicated by the X in the figure). The road
blockage may leave a fuel station inaccessible. 
The other car may therefore alter a scheduled fuel stops and use another fuel station.

\begin{figure}
\centering
\includegraphics[width=0.3\textwidth]{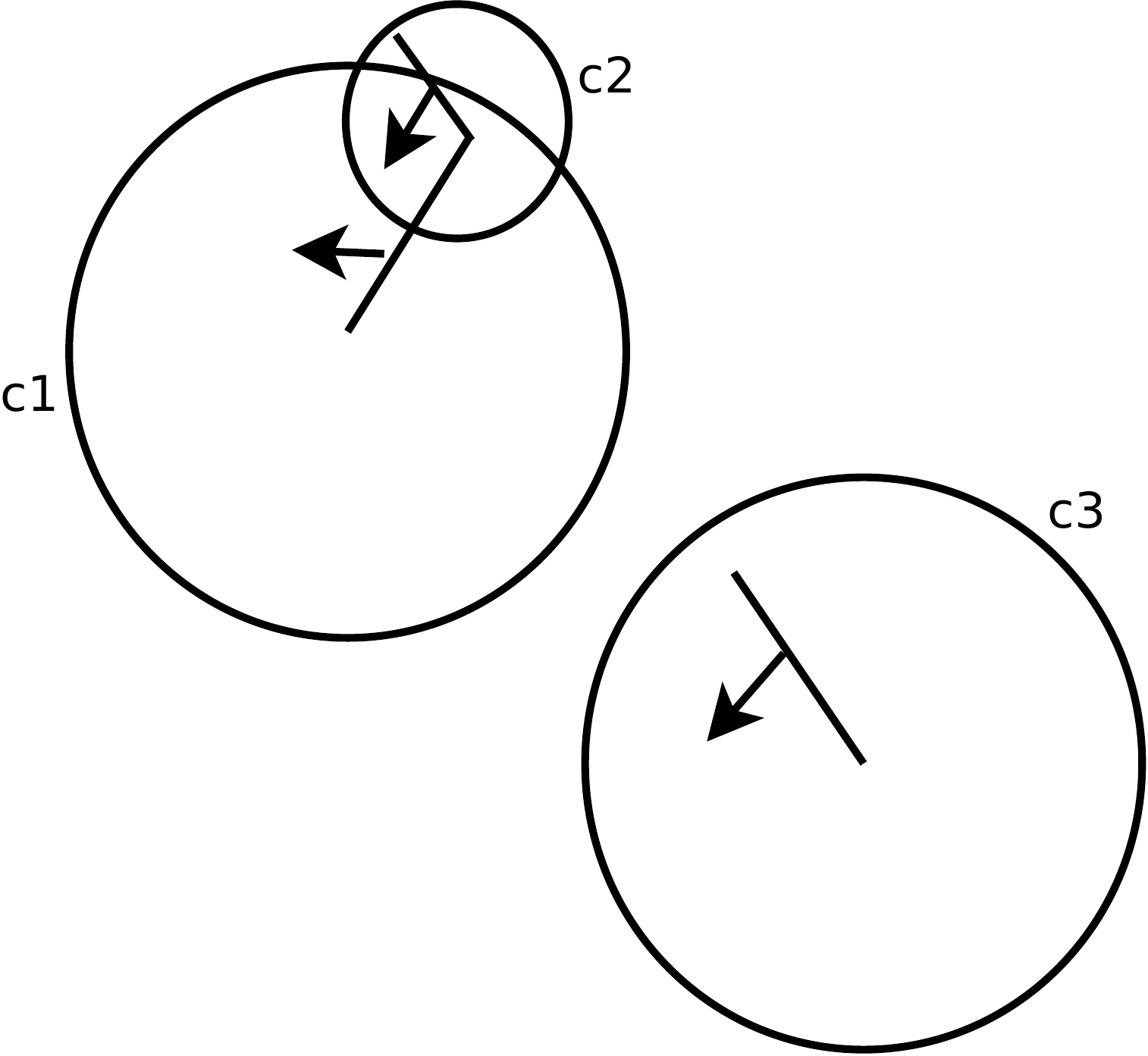}
\caption{Rotating robots schematic system overview}
\label{fig:syss}
\end{figure}

\paragraph{Rotating Robots}
Figure~\ref{fig:syss} depicts an example demonstrating component movement relative to another component. It shows two robots --
components $c1$ and $c3$ are performing circular movements. Robot $c1$ features an attachment
point. In additional component $c2$: a robot arm is attached to this
attachment point. $c2$ also performs circular movements, relative to the
movement of $c1$.

A question one wants to investigate is whether a collision between
$c3$ and $c1$ is possible.

\subsection{System Layers}

We propose the use of four different layers for modeling systems and
their behavior in terms of space, internal behavior and communication.

\begin{enumerate}
\item
A topological or geospatial coordinate system. In our window cleaning
example this is provided by the window itself. In the communicating
cars example this is provided by the topological graph representing
the roads and its geographic / geometric interpretation.
\item
Static components and their interconnections. In the sense of for
example a street network, we can have road blocks and road
construction sites as well as refuel stations.  
In the window cleaning example, we can have single obstacles on the
surface. 
\item
Mobile components. These comprise the cars and robots from our
examples moving within or across their allocated segment without changes in the static components structure. Behavior, may be expressed using geometric expressions. 
\item
Information flow. This can occur between mobile and static components
such as cars or robots or a fuel station.
\end{enumerate}
Systems can evolve and change their structure and behavior.  A new road
may be added or a new fuel station. Cars can move and communicate with
each others. The layers
are ordered with respect to the rate of expected changes, with layer 1 seeing the
least changes and level 4 most changes over time.

\subsection{Specification Layers}
We allow different layers of abstractions for describing systems
and their components. One can mix different specification methods to
describe a widely distributed system, e.g., for covering the
interactions, spatial behavior (topological and geometric) and aspects of state changes within components.

\paragraph{Specification Formalisms}
For describing components and their interactions on a requirements level, the
following possibilities are reasonable:
\begin{itemize}
\item {\it UML state charts}, {\it Message Sequence Charts} and {\it
    Concurrent automata} can be used to specify the behavior of single
  components and their interactions.
\item {\it process algebra terms} can be used as another way to
  specify components and their interactions.
\item We can instrument {\it Pieces of program code} (e.g., a domain specific language, but
  also plain Scala or Java code can be used). The instrumentation can
  be used to generate sequences of behavior at runtime and eventually
  create behavioral descriptions from them. 
\end{itemize}
Furthermore, for describing interactions between components, we use:
\begin{itemize}
\item Logical formula describing events and time points implying interactions.
\item Partially ordered sets of events or shared time points, are used
  to describe shared time where different components may interact with
  each other.
\end{itemize}

\paragraph{Components}
As described above, we distinguish static and mobile components, these
can be aggregated into subcomponents, e.g., an industrial robot can be
made up of arm segments and a mounted tool. Each subcomponent can have
its own behavioral descriptions.    

\paragraph{Time}
Time in our models is defined by a set of time points $T$ and a partial order on
these time points $\lesssim_T$. Each component specification is associated with a set
of time points that is relevant for it. Two time points can form a
time interval. Which can serve as a means for a safe overapproximation of
behavior that is happening in between two time points.

\paragraph{Properties}
In particular we are interested in properties that deal with
overlapping conditions of space.
\begin{itemize}
\item In case of avoiding collisions we are dealing with occupied
  space for components. The occupied space specified in our
  abstractions that are the basis for verification and analysis must
  present an overapproximation of the space that is really used in the
  real-live system to derive a safe (cf. \cite{graf95}) result. We require that
  occupied space of components may overlap at no shared time point or
  time interval. 
\item In case of communication possibilities, with limited ranges, we
  work with space that is a safe underapproximation of a sender and
  receiver range. Here, we can require space inclusion for a set of
  shared time points or time intervals.
\end{itemize} 
We can combine safe overapproximations and underapproximations into
properties, e.g., to state that a component is always within a certain
range of another component, but never comes too close (safety margin)
given certain speed limits.

\section{Reasoning about Spatial Models}
\label{sec:reason}
We describe the verification part of our framework in this section.
\subsection{Overview}
\begin{figure}[htb]
\centering
\includegraphics[width=0.4\textwidth]{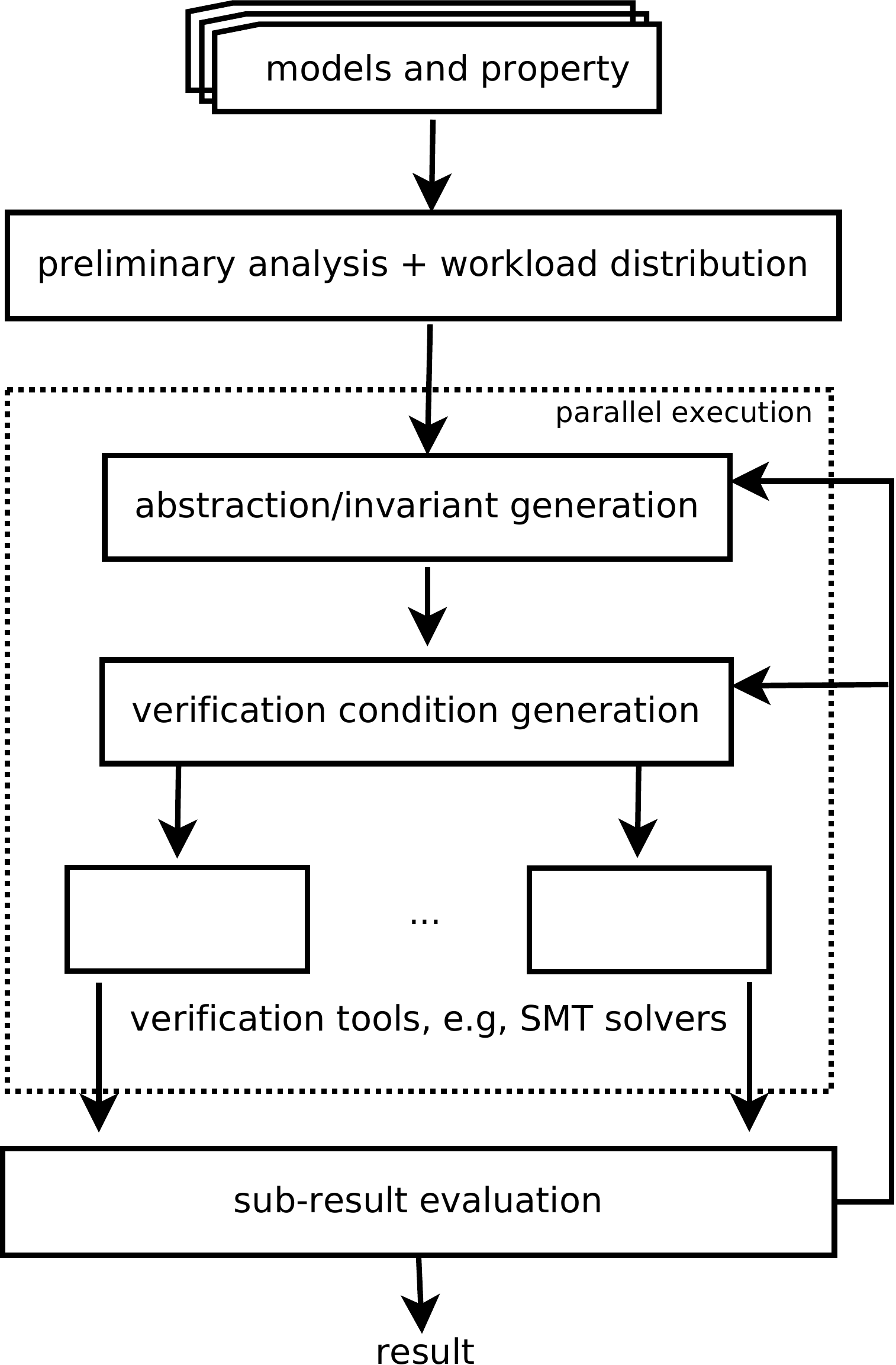}
\caption{Tool workflow}
\label{fig:wf}
\end{figure}

Figure~\ref{fig:wf} shows the workflow for checking the properties of
models with respect to involved computation steps and tools. Models and properties are given to our tool chain for checking. 
\begin{itemize}
\item
In a first step, an algorithm is
used to identify large scale verification goals. For example static components
that are separated in space and do not interact with each other may be
checked independently. The behavior of mobile components may also be
checked independently if they are only acting in a local area in space
or time. Properties may only regard a local area and therefore only the behavior of components that act within this local area has to be taken into account. Best practices for this approach can be scenario and domain specific. For determining whether one component or a property depends on another, techniques from model checking like cone-of-influence reduction \cite{modelch} can be applied. A result of this step is the identification of a set of relevant components for the desired properties.
\item
In a next step, we compute the abstractions of behavior: {\it invariants} for all relevant
components. One way to do this is to unfold all possible execution
traces of an automaton, include all relevant events and annotate them
with time points. These traces are than formalized as invariants.
\item
We use these invariants to generate verification conditions in
a next step. 
Verification conditions can be checked by separate highly
specialised tools like SAT~\footnote{e.g., by using Sat4j:
  \url{http://www.sat4j.org/}} and SMT solvers (e.g., Yices~\cite{yices} or z3~\cite{z3}). 
\item The last step collects the results of used verification tools and
  presents an overall results. Optionally, we refine invariants or
  verification conditions, if the result does not satisfy our
  needs. Eventually this may lead to an iterative process: a
  fixpoint computation or counterexample guided abstraction refinement
  \cite{absref} where invariants are
  the predicates.
\end{itemize}

Computation of invariants can be done in parallel in many
  cases. Verification conditions never depend on each other, so they
  can be always checked in parallel.

\subsection{Component Oriented Workflow}
The checking workflow of possible interference of two components is
done in the following way.
First, a behavioral abstraction: an invariant is created for
each component which is than broken down or combined with information
from another component's invariant to derive verification conditions.
These are given to a solver. Here, invariants contain only
overapproximations of single components and do not take restrictions
due to possible interference with other components into account. Thus,
overapproximations are safe, but may be coarse.

For achieving more fine grained approximations, optionally, we can create logical
conditions capturing the interference of components. These are
invariants, too. Based on them, we can create verification conditions that further
restrain verification conditions derived solely from the invariants
that capture the behavior of single components.

\subsection{Invariants}
Our invariants capture the entire behavior of a component over time or
distinct aspects of interference between components. They
provide a safe overapproximation of behavior.

\begin{figure}
\centering
\includegraphics[width=0.48\textwidth]{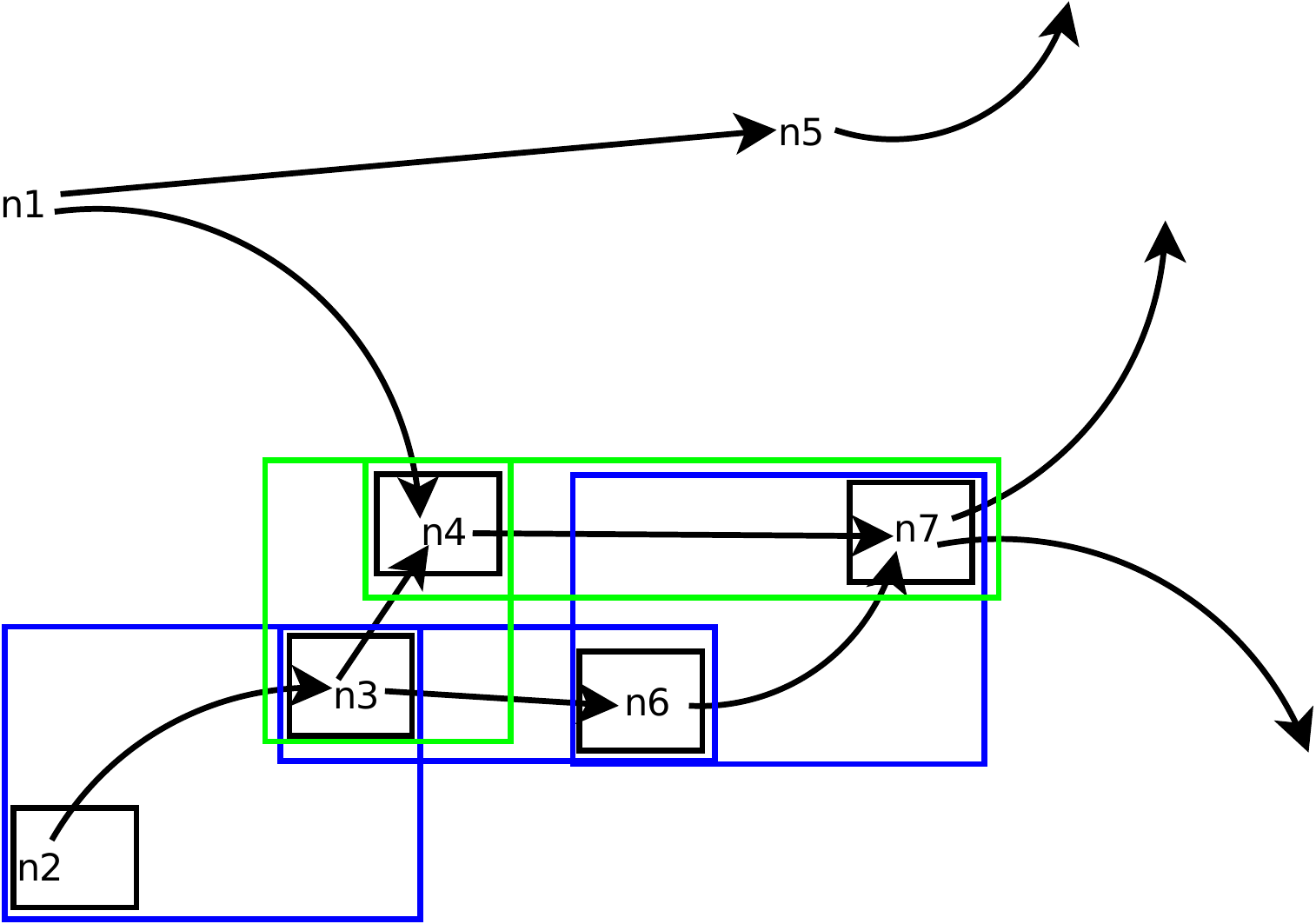}
\caption{Room occupation invariants (time points)}
\label{fig:newexwsinv}
\end{figure}

\paragraph{Nature of Invariants}
Like invariants in classical Hoare style loop verification, our {\it
  spatial component invariants} are supposed to hold over the entire execution time of a
system. Like loop invariants, spatial component invariants typically abstract from the
actual implementation. Similar to loop invariants, where variables' values are often
described in relation to a changing loop index, we describe the
occupied or otherwise classified space and if applicable other
component state based information in relation to time. 
\paragraph{Invariants for Occupied Space}
Figure~\ref{fig:newexwsinv} shows invariants for the second forklift
in our forklift example. These overapproximate the
occupation of space for time points that are associated with the
nodes. Furthermore, overapproximations for time intervals between
time points are shown. In this case the invariants provide a geometric
interpretation of the underlying graph.
Here,  to gain, a safe collision analysis, we must use
overapproximations of the space that is actually occupied.

\begin{figure}
\centering
\includegraphics[width=0.48\textwidth]{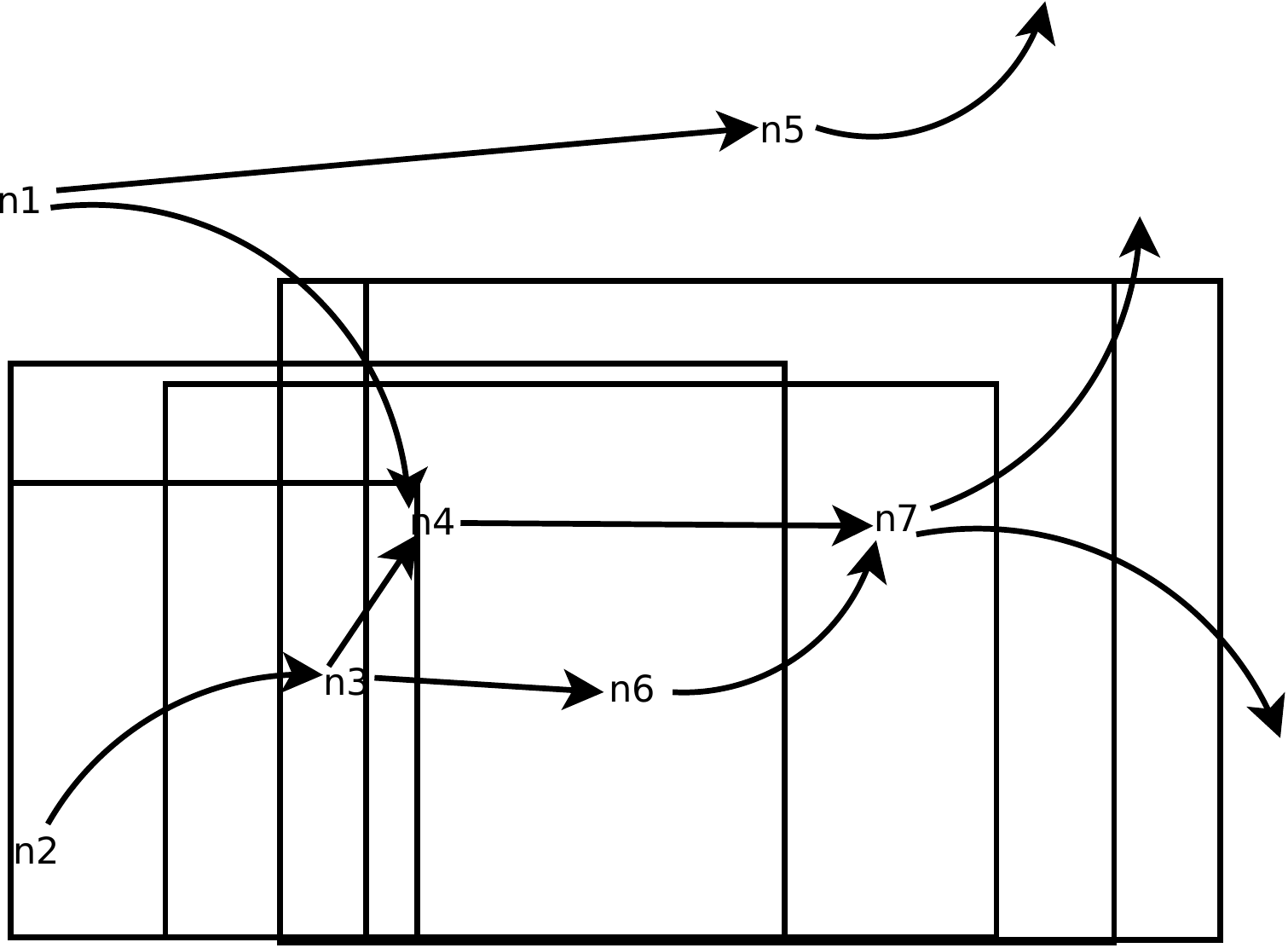}
\caption{Room occupation invariants (time intervals)}
\label{fig:newexwcinv}
\end{figure}

\paragraph{Invariants for Communication and Detection}
Figure~\ref{fig:newexwcinv} shows invariants for the second forklift
in the forklift example indicating visibility
ranges to detect other forklifts. In order to safely analyse detection
possibilities, we need to work with underapproximations. Invariants
characterising time intervals are not shown, but are created from union operations
between the invariants that characterize two neighbouring
time points. 

Note, that for displaying purposes, the granularity of time points in
Figure~\ref{fig:newexwsinv} and Figure~\ref{fig:newexwcinv} is
large. For real systems, the boxes are much closer together, so that
safe 
overapproximations can be much tighter to the actual system behavior. 

\paragraph{Invariants for Interplay between Components}

On an abstract view, we can associate
interaction possibilities between components with events. Events can
also be associated with detection of obstacles or external
interactions. These events can be used in invariants, to
characterize interactions between components.
Such a formula can have the form:
\begin{center}
$($ time point $\wedge$ event $)$ $\longrightarrow$ occupied space \&
other actions
\end{center}
We can create invariants that describe aspects of interactions
between components.  Some possible conditions on triggering events can have the
  forms
\begin{center}
time point $\rightarrow$ event \\
or\\
$($ time point $\wedge$ event $)$ $\rightarrow$ event' \\
\end{center}
Combining these invariants can be done by intersection with other
invariants characterizing the behavior of a single component in order
to restrain its behavior and derive more refined verification conditions
that are given to solvers.

\begin{figure}
\centering 
\includegraphics[width=0.4\textwidth]{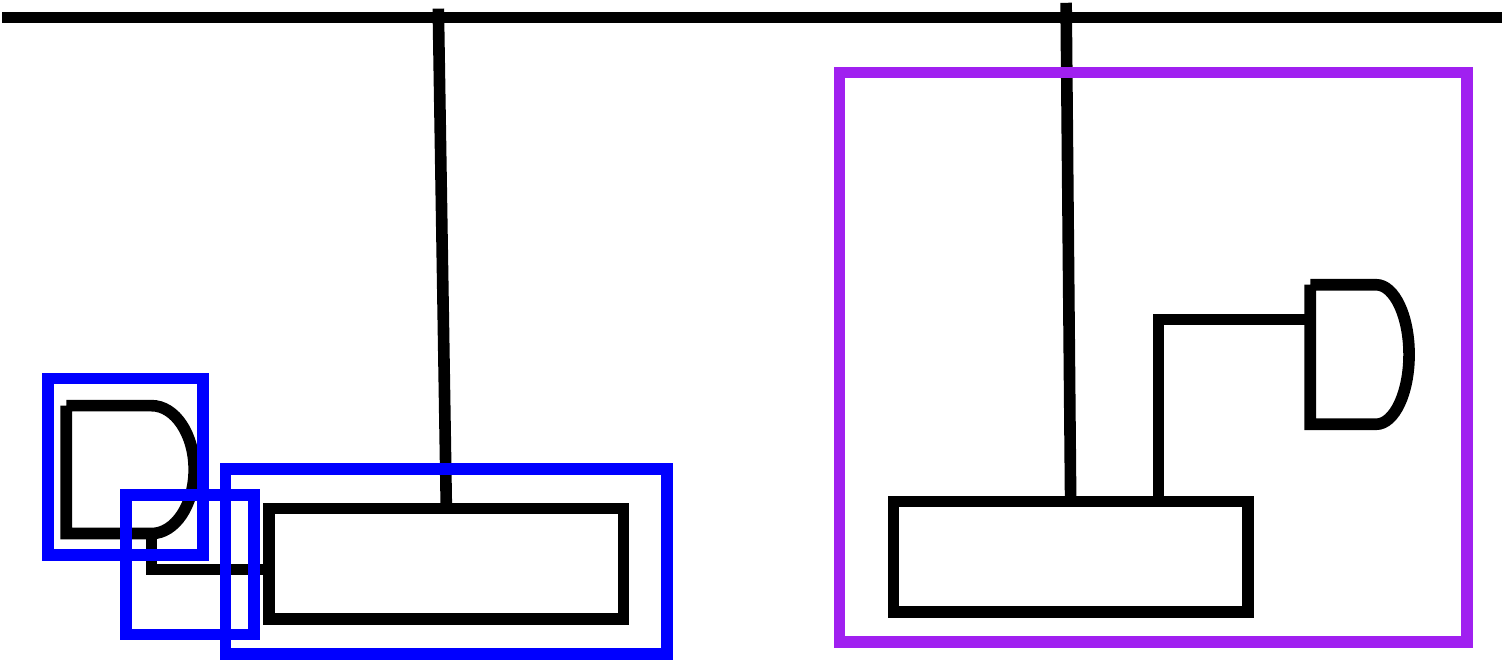}
\caption{Concurrent window cleaning}
\label{fig:ex1a}
\end{figure}

\paragraph{Aggregating Subcomponents}
We can aggregate subcomponents into larger components. Likewise, we
can integrate their invariants. Figure~\ref{fig:ex1a} shows invariants
characterizing occupied space for a given point in time
for the cleaning platforms from Figure~\ref{fig:ex1}. The right side displays
invariants for subcomponents like the actual platform, the robot arm
and the tool while the left side displays a safe overapproximation of
the combined system.

\paragraph{Term based Representations of Invariants}
We represent invariants syntactically as terms. Invariants are built from
 predicates that are combined using function symbols like
logical operators. An invariant that characerizes the spatial
behavior of a single component has the form: 
\begin{center}
...\\
$ t= i \ \ \ \ \  \rightarrow \mathit{OccupySpace}_i \ \ \ \wedge
\mathit{CommunicationRange}_i \ \ \ \wedge ... $ \\
$ t= i +1 \rightarrow \mathit{OccupySpace}_{i+1} \wedge
\mathit{CommunicationRange}_{i+1} \wedge ...$ \\
$ t= i + 2 \rightarrow \mathit{OccupySpace}_{i+2} \wedge
\mathit{CommunicationRange}_{i+2} \wedge ...$ \\
... \\
\end{center}
For non linear time, the $i$ index is replaced by a more complex time
point indicating the node in the graph characterizing the partial
order of time points. 
We can perform operations on invariants like the merging of time
points into time intervals thereby transforming it into another
syntactic representation. For the example above this is realized as follows:
\begin{center}
...\\ 
$ t \in [i,i+1) \rightarrow \ \ \ \ \ \ \ \ \ \ \ \ \ \ \ \ \ \ \ \ \
\ \ \ \ \ \ \ \ \ \ \ \ \ \ \ \ \ \ \ \ \ \ \ \ \ \ \ \ \ \ \ \ \ \ \
\ $ \\ 
$\mathit{OccupySpace}_i \sqcup
\mathit{OccupySpace}_{i+1}  \wedge \ \ \ $ \\
$ \mathit{CommunicationRange}_i  \ \ \sqcup $ 
$\mathit{CommunicationRange}_{i+1}  \wedge ... $ \\ 
$ t \in [i+1,i+2) \rightarrow \ \ \ \ \ \ \ \ \ \ \ \ \ \ \ \ \ \ \ \
\ \ \ \ \ \ \ \ \ \ \ \ \ \ \ \ \ \ \ \ \ \ \ \ \ \ \ \ \ \ \ \ $ \\ 
$\mathit{OccupySpace}_{i+1} \sqcup 
 \mathit{OccupySpace}_{i+2} \wedge \ \ \ $ \\
$ \mathit{CommunicationRange}_{i+1} \ \ \sqcup $ 
$\mathit{CommunicationRange}_{i+2}  \wedge ... $ \\ 
...
\end{center}
Here, $\rightarrow$ and $\wedge$ are function symbols of the term. The
$\sqcup $ is a function that performs an interpretation and
yields a subterm representing a semantic union of its arguments.

Furthermore, we have operations that comprise simplifications, abstractions,
normalizations and assigning and removing ownership -- means to
indicate that space belongs to a distinct component  -- to occupied
space. Contrary to abstractions we have interpretations on the invariant level, e.g., for assigning a
geometry to parts of topological or high-level component representations, thereby lifting
invariants to a more concrete level.

\paragraph{Generating Invariants}
Invariants may be directly encoded or generated in the following ways from spatial behavioral specifications.
\begin{itemize}
\item {\it Unfolding of automata to generate timed conditions in
  invariants.} If we are only interested in checking a limited time
  span, this is straight forward. For periodically repeating behavior,
  we can group time points into equivalence classes for a finite
  invariant that describes the infinite behavior. A special case of
  automata unfolding are sequences of states that a system component
  has gone through during a simulation. These sequences can be
  generated from other tools such \footnote{We have successfully applied it
  together with Reactive Blocks and its specification method \cite{arctis}}.
\item {\it Instrumentation of code pieces to generate timed conditions in
  invariants at runtime during a testing phase of a system.} Here, we
  have to ensure that we cover all relevant aspects of a system, because we
  are in danger of getting an unsafe underapproximation of the
  behavior. However, regarding the behavior of robots, it seems
  reasonable to exhaustively catch all spatial movements of a robot
  without having exhausted the robot's software state space. 
\item {\it Transformation of specifications on a symbolical /
    semantical level.} This involves symbolic analysis of the program
  code that controls a distributed system and the extraction of
  invariants based on them. 
\end{itemize}

\begin{figure}
\centering
\includegraphics[width=0.46\textwidth]{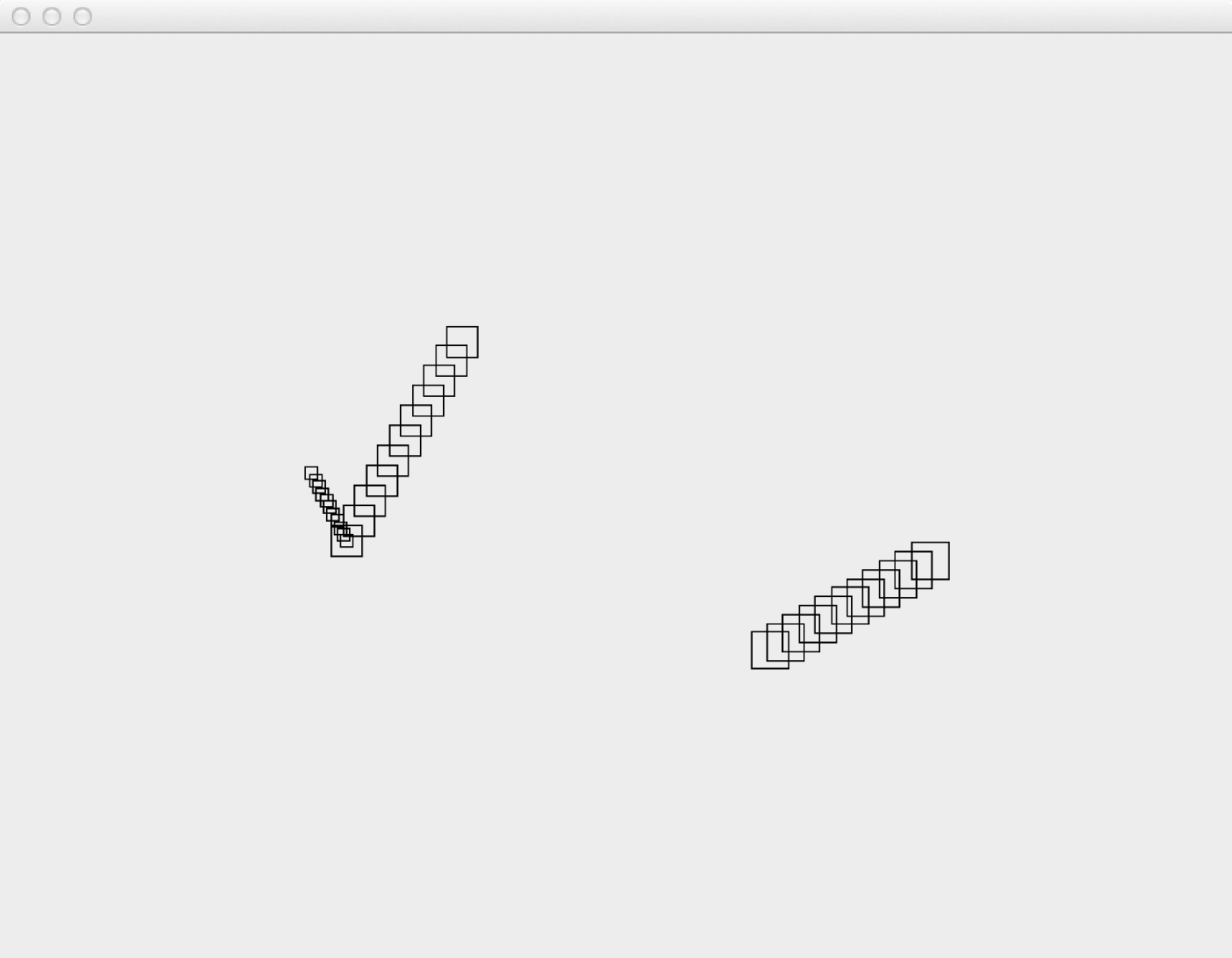}
\includegraphics[width=0.46\textwidth]{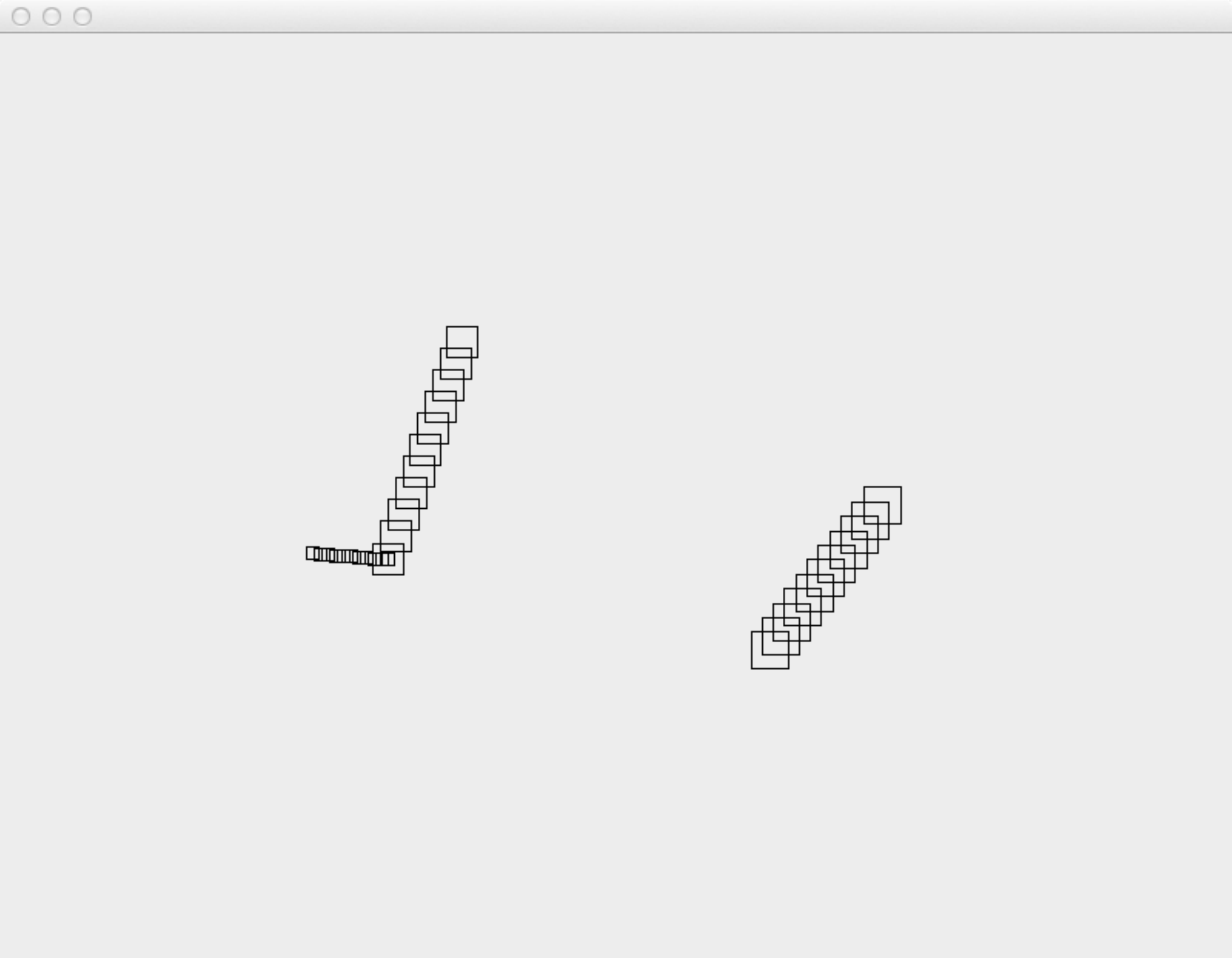}
\includegraphics[width=0.46\textwidth]{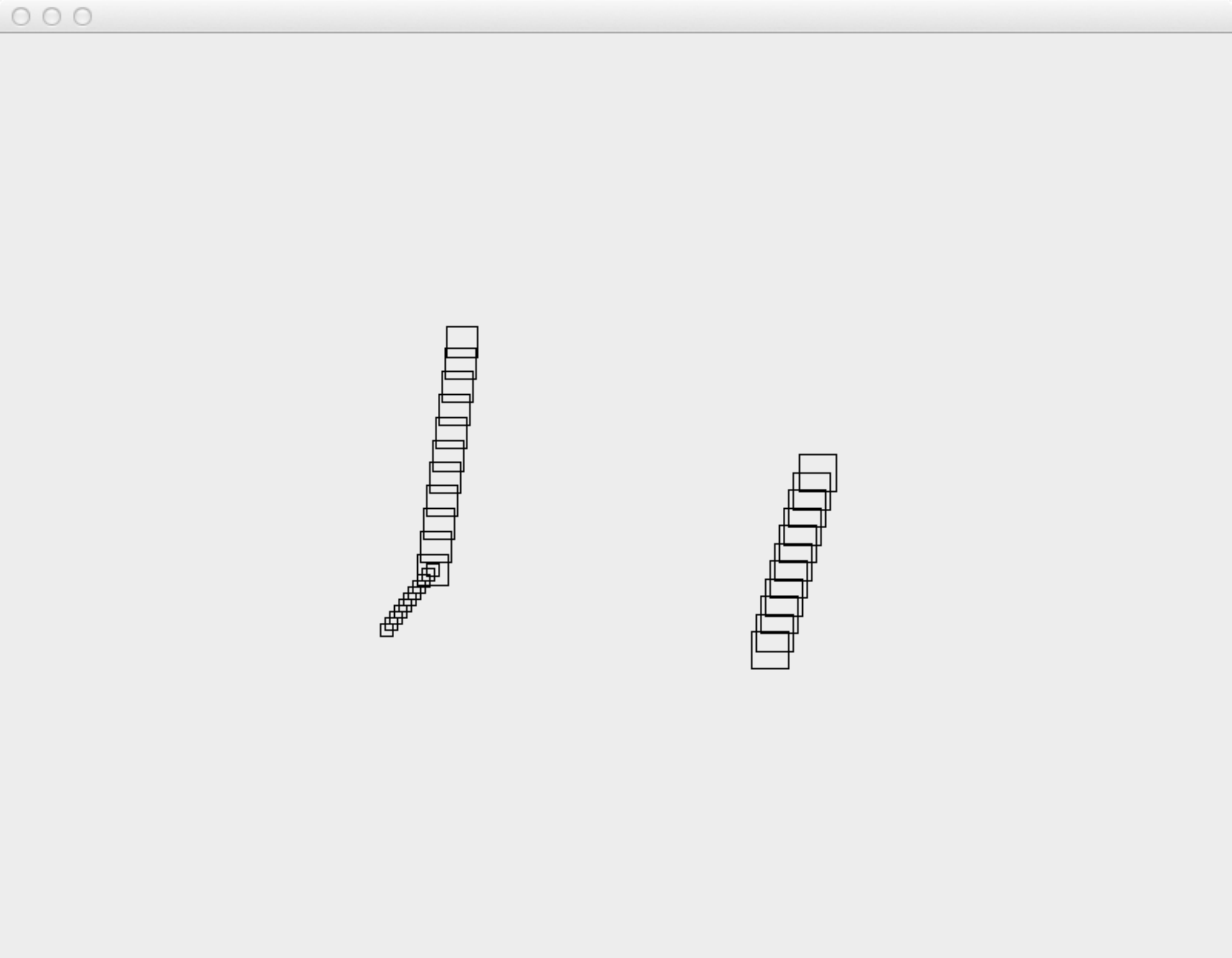}
\includegraphics[width=0.46\textwidth]{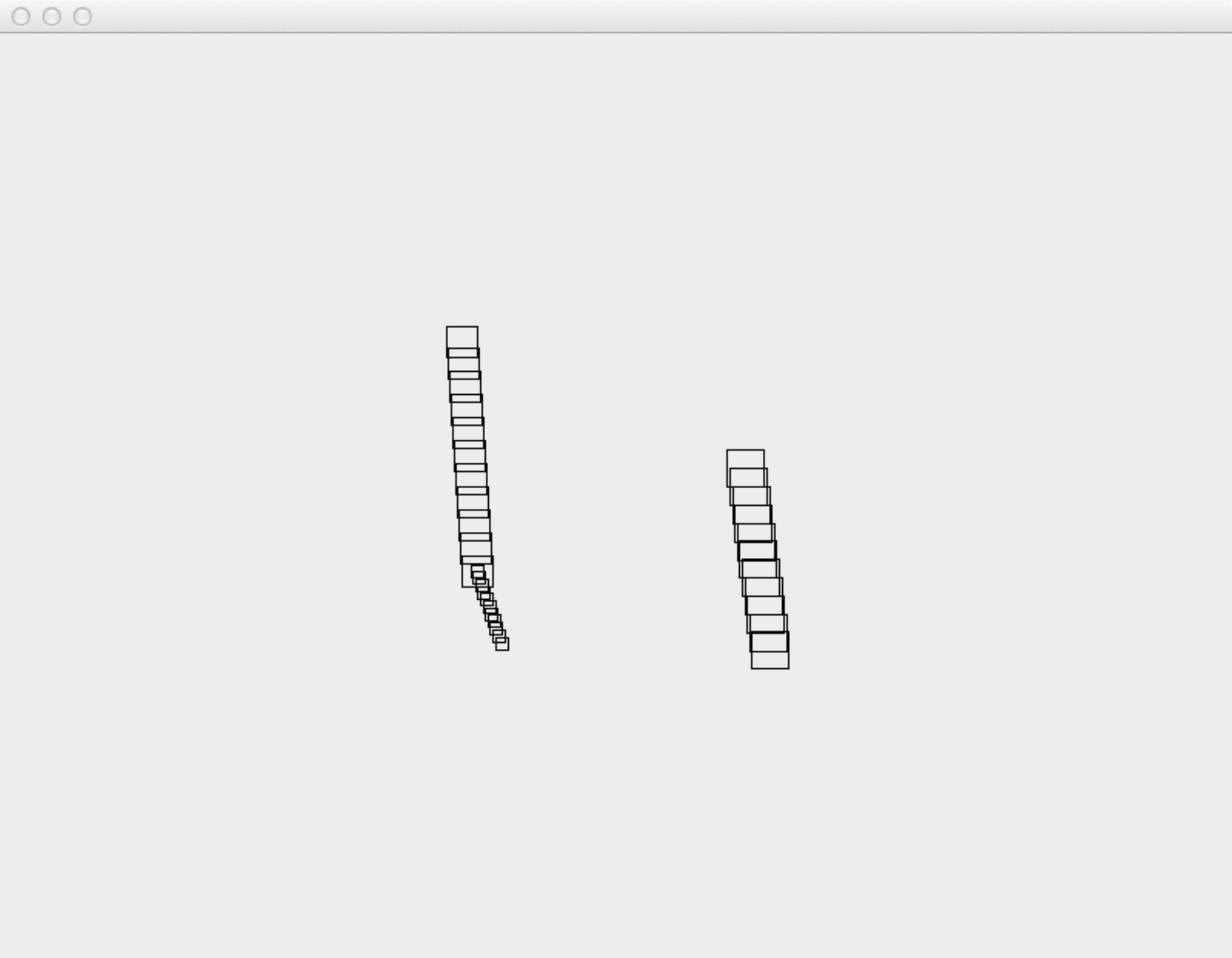}
\includegraphics[width=0.46\textwidth]{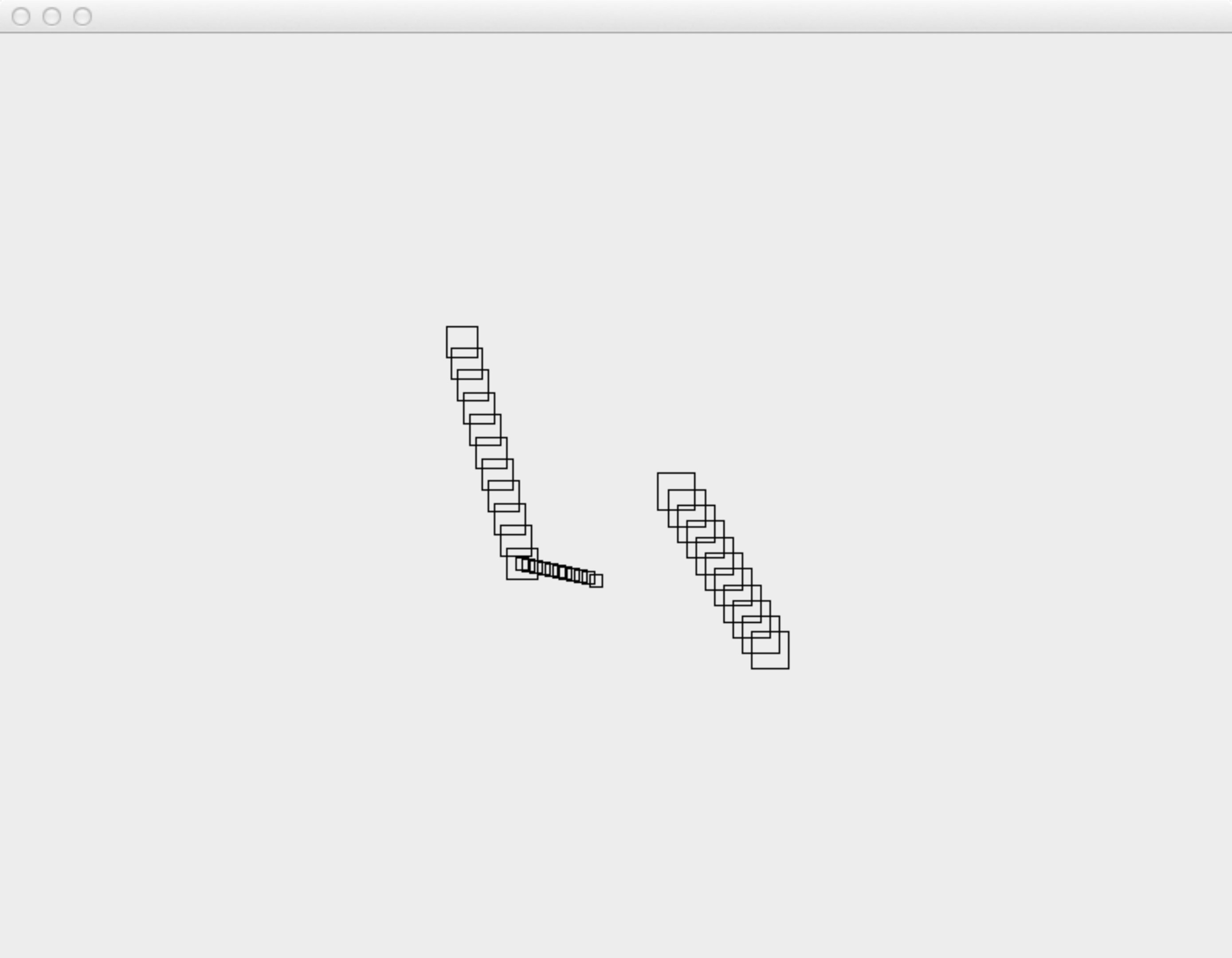}
\includegraphics[width=0.46\textwidth]{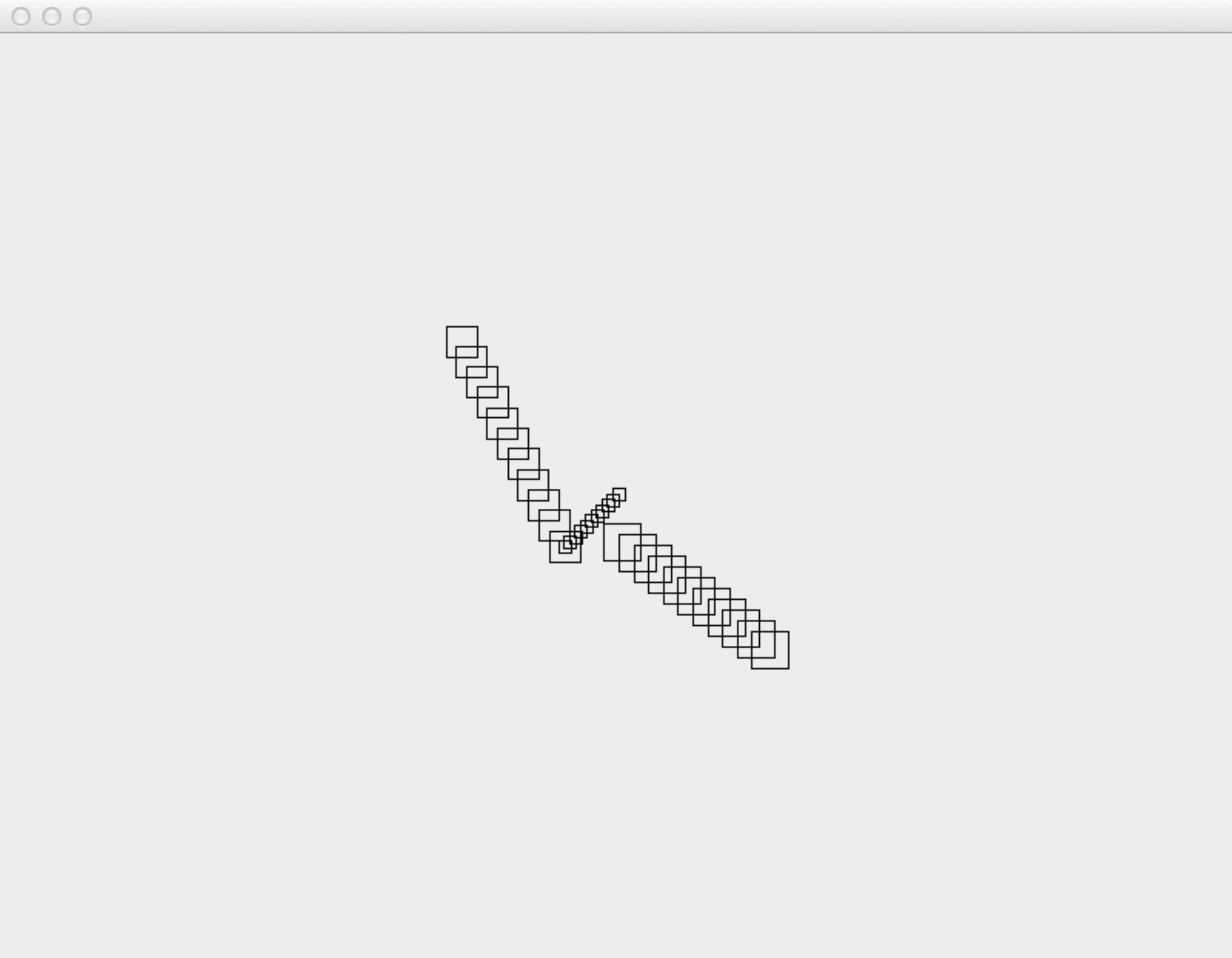}
\caption{Box-based overapproximation (Screenshots)}
\label{fig:bbo}
\end{figure}

\paragraph{Abstractions using Invariants}
We can use abstraction on the invariant level to facilitate later
verification tasks. For space, we can use boxes to over- or
underapproximate more detailed structures.
Figure~\ref{fig:bbo} shows a screenshot of a box based overapproximation in the
rotating robots example for collision detection. 
Here, the exact amount of space for segments of robot components is automatically
abstracted using overapproximations.

\subsection{Verification Conditions}
We connect different tools including SAT and SMT solvers to our
framework. Verification conditions are generated from invariants so
that a property can be verified and the input is suitable for the
particular tool.
\paragraph{Predicate Based Space Conditions}
In our invariants we can use predicates to characterize the occupation of space for a
distinct time point or time interval. As an example: each point indicating the
occupation of a square on a plane can be identified using a
predicate. By doing so, space occupied by different components can be
made comparable on the invariant level. Shown in Figure~\ref{fig:predunf} is the unfolding of a predicate characterizing
the occupation of a box specified by an upper left  and a lower right
pair of coordinates.
\begin{figure*}
{\footnotesize\parindent1.2cm
\indent $ occupyXYspace (x_1,y_1,x_2,y_2) = $ \\
\indent\indent $occupyXY(x_1,y_1) \wedge occupyXY(x_1+0.1,y_1) \wedge ... \wedge occupyXY(x_2,y_1) \wedge $ \\ 
\indent\indent $ occupyXY(x_1,y_1+0.1) \wedge occupyXY(x_1+0.1,y_1+0.1) \wedge ... \wedge occupyXY(x_2,y_1+0.1) \wedge $ \\
\indent\indent\indent $...$ \\
\indent\indent $ occupyXY(x_1,y_2 - 0.1) \wedge occupyXY(x_1+0.1,y_2-0.1) \wedge ... \wedge occupyXY(x_2,y_2-0.1) \wedge $ \\
\indent\indent $ occupyXY(x_1,y_2) \wedge occupyXY(x_1+0.1,y_2) \wedge ... \wedge occupyXY(x_2,y_2)$ }
\caption{Predicate unfolding (conjunction of points)}
\label{fig:predunf}
\end{figure*}
The resulting representation can be compared with
unfolded invariants of other components. This comparison must be
made for each shared time point or time interval between components and can by
either given to a SAT solvers or can by realized by a small algorithm
implemented in Scala in our framework.

\paragraph{Space Conditions Based on Inequalities}

 A second way of comparing different invariants that characterize
 geometric occupation of space is to generate for each shared
  time point a geometric formula characterizing the potential
  overlapping of occupied space. For example,  two components
  $c_1$ and $c_2$
  specified with a spatial occupation specified by boxes with
  upper left and lower right
  coordinates as given below: 
\begin{center}
  $(x_{1_{c}}(t),y_{1_{c}}(t))$,  $(x_{2_{c}}(t),y_{1_{c}}(t))$, \\
          $(x_{1_{c}}(t),y_{2_{c}}(t))$, $(x_{2_{c}}(t),y_{2_{c}}(t))$ 
\end{center}
Where each component is given by a function that depends on a time
parameter $t$ we generate the following:
\begin{center}
$\neg \exists x, y. $ \ \
$\bigvee_{t \in \mathit{shared \ time \ points}} $ \\
$(x_{1_{c_1}}(t) \le x \le x_{2_{c_1}}(t) \wedge$ \\
$x_{1_{c_2}}(t) \le x \le x_{2_{c_2}}(t) \wedge$ \\
$y_{1_{c_2}}(t) \le y \le y_{2_{c_2}}(t) \wedge$ \\
$y_{1_{c_2}}(t) \le y \le y_{2_{c_2}}(t)  )$
\end{center}
A solution characterizes the overlapping at any shared time point. If
no solution exists, there is no overlapping. We can use this  as
a verification condition thereby unfolding the $\bigvee_{t \in
  \mathit{shared \ time points}} $ and treating multiple time points or
generate an independent verification condition for each time point.

\section{Implementation and Application}
\label{sec:impl}
Here, we describe some features of our framework that we have
implemented and evaluated.
\subsection{Specification of Models and Properties}
Invariants and operations on them are encoded using abstract data type style {\it
  case classes} from the Scala language. 
\paragraph{Basic Constructs}
To give a look and feel, a small excerpt of the abstract datatypes in
Scala is given below:
{\footnotesize
\begin{verbatim}
abstract class Invariant;

abstract class ATOM extends Invariant;

case class OR (t1 : Invariant, t2 : Invariant) 
   extends Invariant;
case class AND (t1 : Invariant, t2 : Invariant)  
   extends Invariant;
case class NOT (t : Invariant) extends Invariant;
case class IMPLIES (t1 : Invariant, t2 : Invariant)  
   extends Invariant;
        ...
case class BIGOR (t : List[Invariant]) extends Invariant;
case class BIGAND (t : List[Invariant]) extends Invariant;
        ...
case class TimePoint [T](timepoint : T) 
   extends ATOM; 
case class TimeInterval [T]
 (timepoint1 : T, timepoint2 : T) 
   extends ATOM; 
case class Event[E] (event : E) extends ATOM;
        ...
case class Occupy3DBox 
  (x1 : Int, y1: Int, z1 : Int, 
   x2 : Int, y2 : Int, z2 : Int) extends ATOM;
case class OccupySegment3D 
  (x1 : Int, y1 : Int, z1 : Int,
   x2 : Int, y2 :Int, z2 : Int, radius : Int) 
     extends ATOM;
case class Occupy3DPoint (x:Int, y:Int, z: Int) 
     extends ATOM
\end{verbatim}}
The excerpt shows only some logical operations, time (the actual type
used for representing time is parameterized {\tt T}, any type with a partial
order is suitable) and geometric
constructs for an Euclidian space. Some abbreviations for different
levels of modeling and easier automatic processing can be seen. For example, the {\tt BIGOR} and {\tt
  BIGAND} constructs are semantically equivalent to nested {\tt AND}
and {\tt OR} constructs.
\paragraph{Ownership of Geometric Structures}
We support the ownership of geometric structures to identify
structures that belong to a specific aspect of an entity. For example,
the following constructors are available:
{\footnotesize
\begin{verbatim}
case class OwnBox[C] (owningcomponent : C,x1 : Int,y1 : Int,x2 :
Int,y2 : Int) extends ATOM; 
// a box that is owned by a component.
        ...
case class OwnPoint[C] (owningcomponent : C,x:Int, y:Int) extends ATOM
\end{verbatim}}
\paragraph{Relative Time Points and Symbolic Integer Values}
In addition to absolute time, we also allow the specification of time
relative to an event (event relative time point {\tt ERTP}) and the use of symbolic values as integers for
coordinates (type {\tt SI}. Geometric constructs making use of these
types are provided. For example, the following constructors are
available:
{\footnotesize
\begin{verbatim}
abstract class TIMEPOINT;

abstract class ERTP; // Event Relative Time Point

case class TERTP [E,T] (event : E, offset : T) extends ERTP;
case class IntERTP [E](event : E, offset : Int) extends ERTP; 

abstract class SI; // SymbolicInt
case class SI_C (c:Int) extends SI;
case class SI_Add (i1 : SI, i2 : SI) extends SI;
case class SI_Sub (i1 : SI, i2 : SI) extends SI;
case class SI_Times (i1 : SI, i2 :SI) extends SI;
case class SI_Var [V](v : V) extends SI
        ...
case class OccupyBoxSI (x1 : SI,y1 : SI,x2 : SI,y2 : SI) extends ATOM; 
// symbolic coordinates, alternative concept to EROccupyBox
case class EROccupyBox (
   x1 : (ERTP => Int),y1 : (ERTP => Int),
   x2 : (ERTP => Int),y2 : (ERTP => Int)) extends ATOM; 
// Coordinates depending on an event relative time point
\end{verbatim}}

\paragraph{Specification Example}
A very small example for an invariant encoded in Scala describing only the possible movement of a forklift in the graph shown
in Figure~\ref{fig:newex}
is presented below (here, some more abstract representation {\tt
  OccupyNode} is chosen):
{\footnotesize
\begin{verbatim}
  val topologicalinvariant_fl2 : Invariant = 
   AND( AND ( AND (
      IMPLIES(TimePoint("pt1"),OccupyNode("n2")), 
      IMPLIES(TimePoint("pt2"),
         OR (OccupyNode("n3"),OccupyNode("n4")))), 
      IMPLIES(TimePoint("pt3"),
         OR (OccupyNode("n6"),OccupyNode("n7")))),
      IMPLIES(TimePoint("pt4"),OccupyNode("n7"))); 
\end{verbatim}}
Only 4 different time points and no geometric interpretation
are encoded.

In realistic examples, we use functions that build abstract datatypes
comprising thousands of constructors based. The construction can be
based on parameters to
characterize different behaviors and interaction possibilities of
different components. 

\subsection{Parameterized Specifications}
Parameterized specifications can be created by Scala functions that
create distinct formalizations based on their parameters.

\paragraph{Lifting of a Workpiece}

\begin{figure}
\centering
\includegraphics[width=0.48\textwidth]{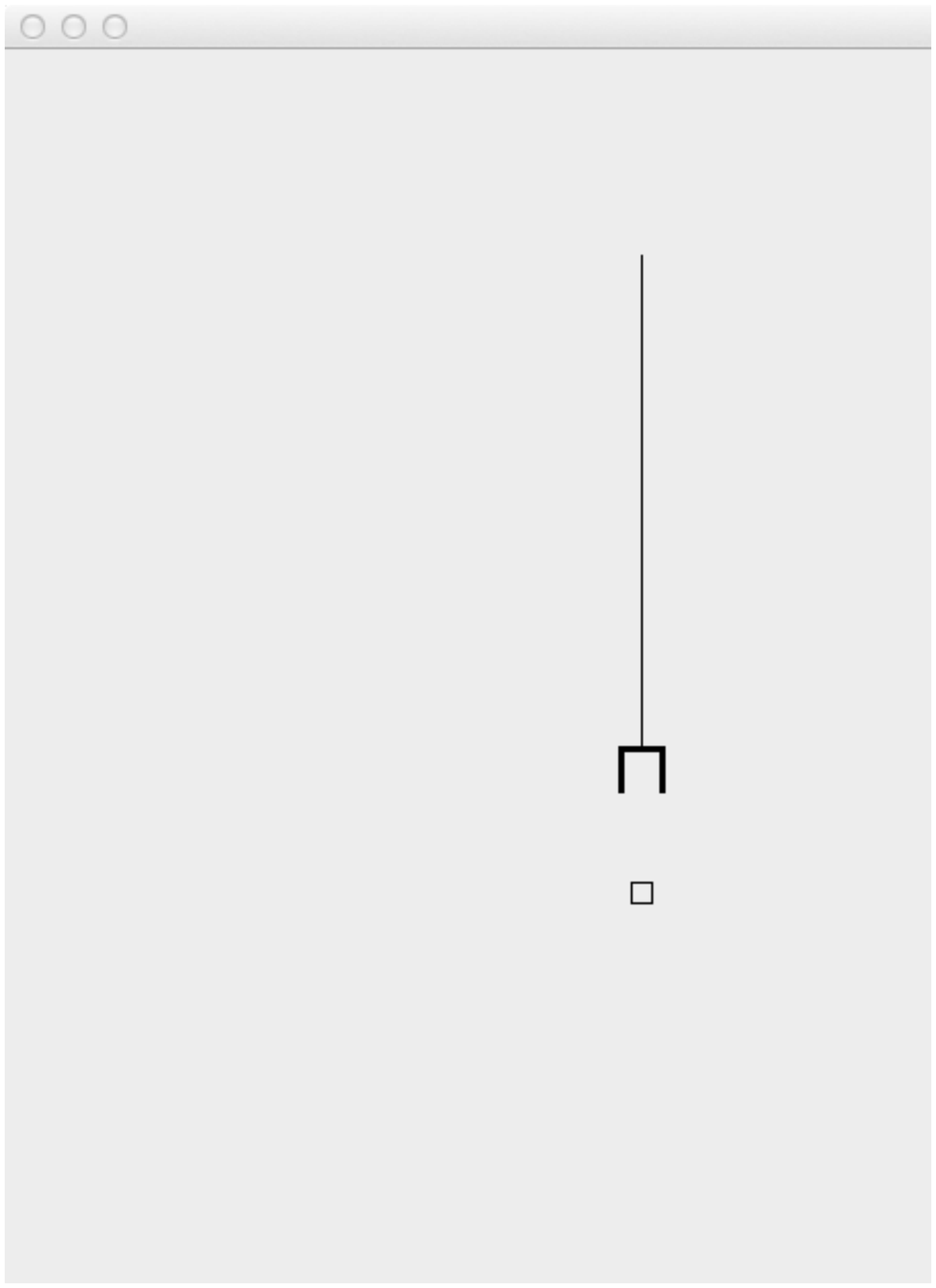}
\includegraphics[width=0.48\textwidth]{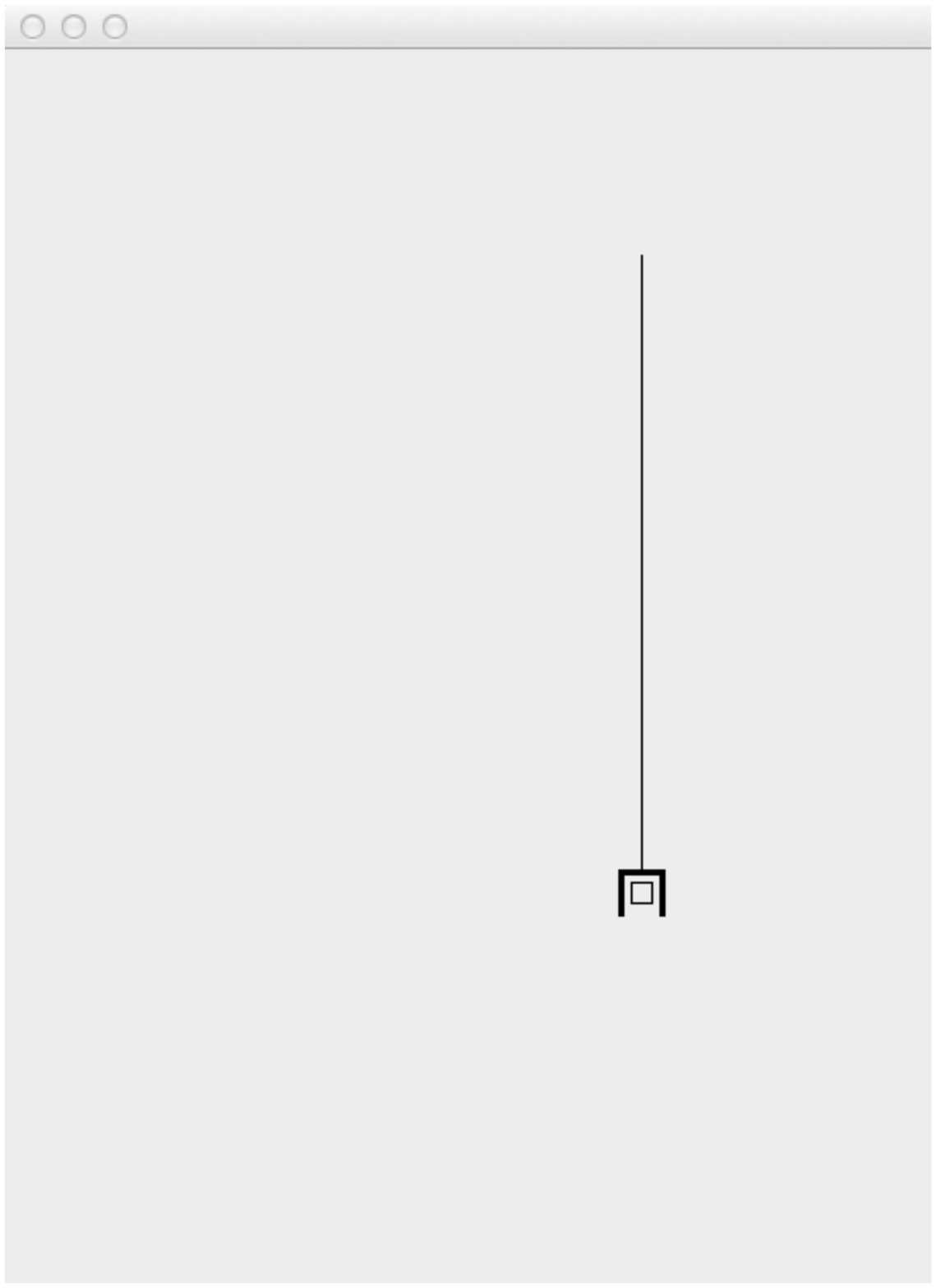}
\includegraphics[width=0.48\textwidth]{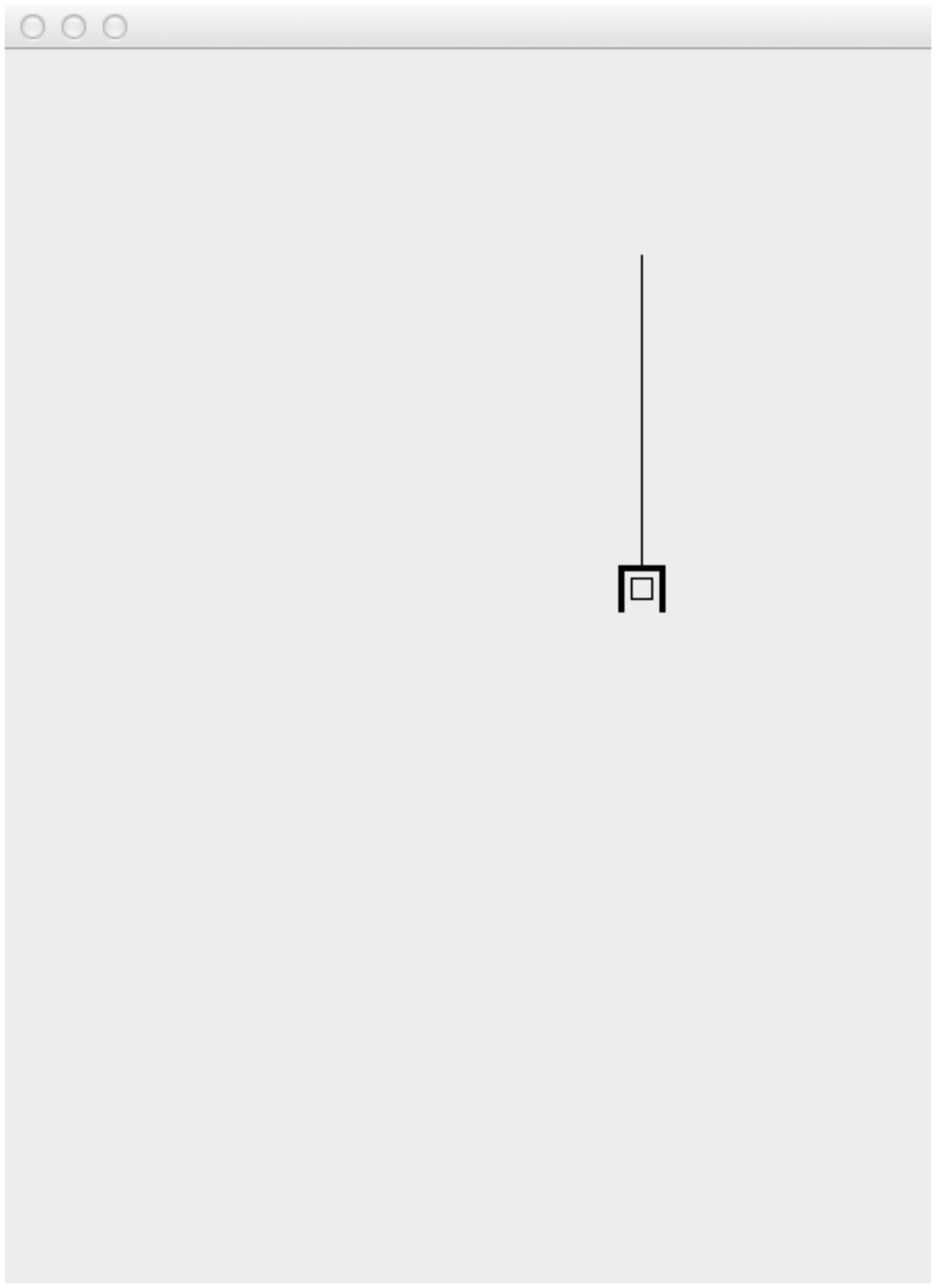}

\caption{Workpiece lifting}
\label{fig:wplift}
\end{figure}

Figure~\ref{fig:wplift} shows a grapple hook lifting a workpiece. The
behavior of the hook and the workpiece is formalized in BeSL.
Here, a grapple hook consists of four subcomponents: the two arms for
holding the workpiece on the sides, a components that links the two
arms and a chain that can be lowered or raised to position the
hook. The workpiece is just formalized as a single component.

\paragraph{Parameterized Lifting Invariants}
We have experimented with different behavioral descriptions for the
hook. One parameterized description for one of the arms is given
below. It formalizes the behavior of for 200 time points. The speed
can be given as a parameter. Furthermore, the lifting may be stopped
for a given point in time.
{\footnotesize
\begin{verbatim}
def invariantR3(speed : Float, stoppointup: Int) : Invariant ={
	var inv : List[Invariant] = Nil;
	  
	for (i <- 0 to 100) {
	  inv ::= (IMPLIES(TimeStamp (i), 
	      OccupySegment(300,200+(i*speed).toInt,320,200+(i*speed).toInt,3)))	
	}
	for (i <- 0 to 100) {
	  	  if (i < stoppointup) {
	  		  inv ::= (IMPLIES(TimeStamp (i+101), 
	  				  OccupySegment(
                     300,
                     200+(100*speed).toInt-(i*speed).toInt,
                     320,
                     200+(100*speed).toInt-(i*speed).toInt,3)))  	    
	  	  } else {
	  		  inv ::= (IMPLIES(TimeStamp (i+101), 
	  				  OccupySegment(
                     300,
                     200+(100*speed).toInt-(stoppointup*speed).toInt,
                     320,
                     200+(100*speed).toInt-(stoppointup*speed).toInt,3)))  	  	    
	  	  }
	}
	return (BIGAND(inv));
}
\end{verbatim}}

\begin{figure}
\centering
\includegraphics[width=0.48\textwidth]{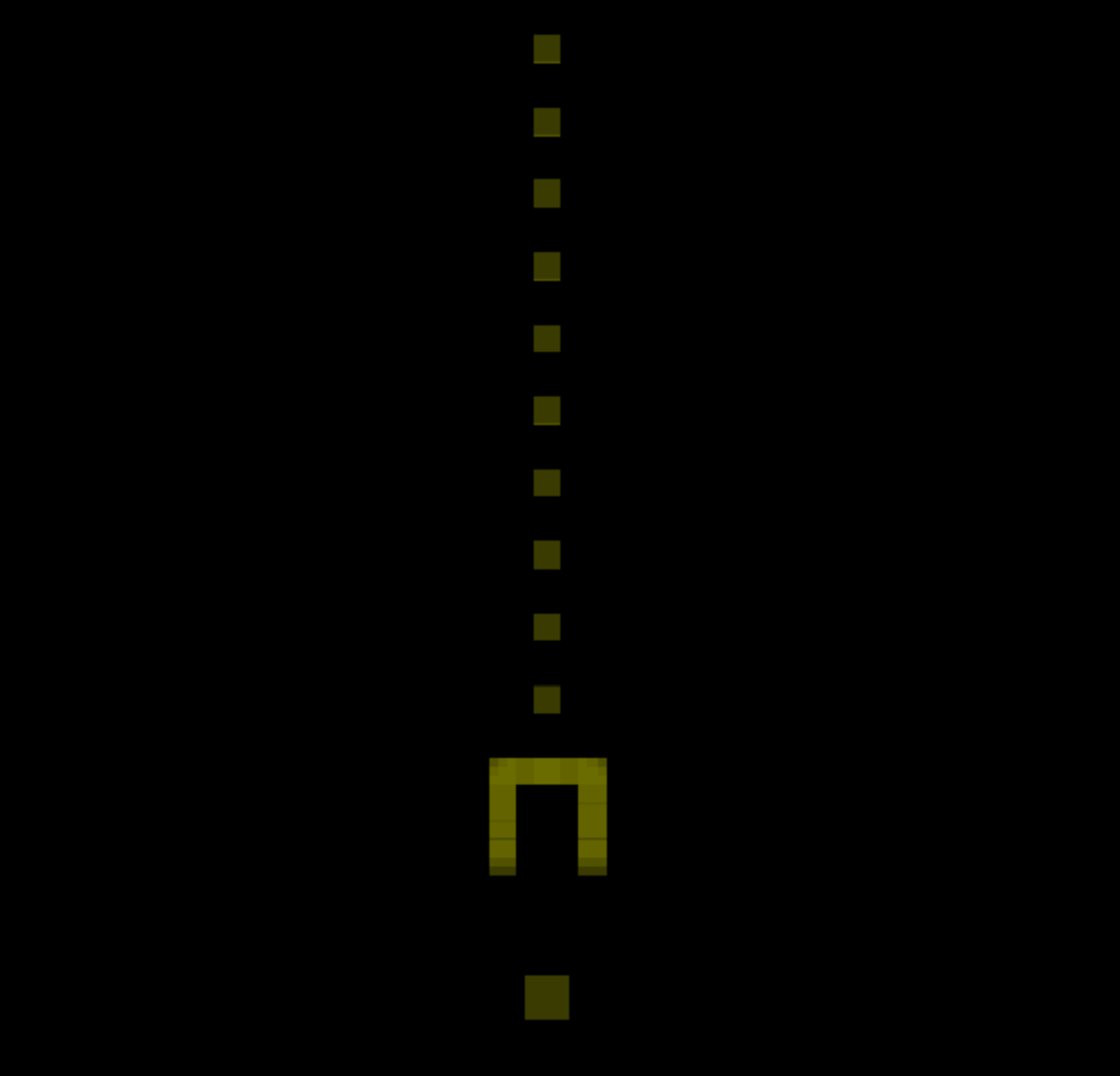}

\caption{Workpiece lifting 3D}
\label{fig:grab3d}
\end{figure}

A 3D version of the workpiece lifting is shown in Figure~\ref{fig:grab3d}. It
contains an abstraction of segment descriptions based on 3D boxes.

\subsection{Checking Collision Conditions}
We have implemented the checking of component invariants for possible
collisions. As a more realistic example and means for measuring performance we have regarded two invariants for collisions where each
invariant comprises 1000 different time point entries and associated geometric
occupied space.\footnote{The following checking times where retrieved for an intel core i5 running 2.8
GHz with 8 GB RAM running Mac OS 10.8.4. The SMT solver used is z3 \cite{z3}.}
\begin{enumerate}
\item Breaking the invariants down to SMT verification conditions
  where one verification condition is generated per time point and
  calling the SMT solver takes between 20 and 25 seconds.
\item Breaking the invariants down to a single SMT verification
  condition and checking can be performed in slightly less then one
  second.
\item Breaking the invariants down to verification conditions each one
  comprising 15000 predicates indicating the occupation of single
  points for one invariant and 20000 predicates indicating the
  occupation of single points for the second invariant for each time
  point can be done in at most 7-8 seconds. Here, a HashSet based Java
  implementation is used. In case of early collision detection the
  checking time can be significantly less. 
\end{enumerate}
From experiments 1 and 2 one can be seen that it is usually a good idea to encode as many
information as possible into an SMT formula, since the overhead of
calling it is significant. 

If higher resolutions are used, the single points' abstractions in the
third experiment
performance decreases linear with the amount of points used. The third experiment can be
  done using a SAT solver, this takes significantly longer, due to the
  higher complexity of SAT formula compared to the Java HashSet and
  the less abstraction power compared to the SMT based approach. 

Collisions have to be checked between each two components for each
shared time point or
time interval. Our notion of time allows the characterization of components
that will never interfere, so the amount of components that one has to
check for a time point does depend on the application
scenario. Analysing whether a time point or time interval is shared or not
can be based on requirements and specification documents. To achieve a
safe result, we must use an overapproximation of shared time points
and time intervals.

Furthermore, it is possible that components have a
non-deterministic behavior. In that case, we need to adapt out SMT
conditions for the experiments one and two or use the approach based
on the experiment three and add all possible occupied point to the HashSet.

\begin{figure}[htb]
\centering
\includegraphics[width=0.3\textwidth]{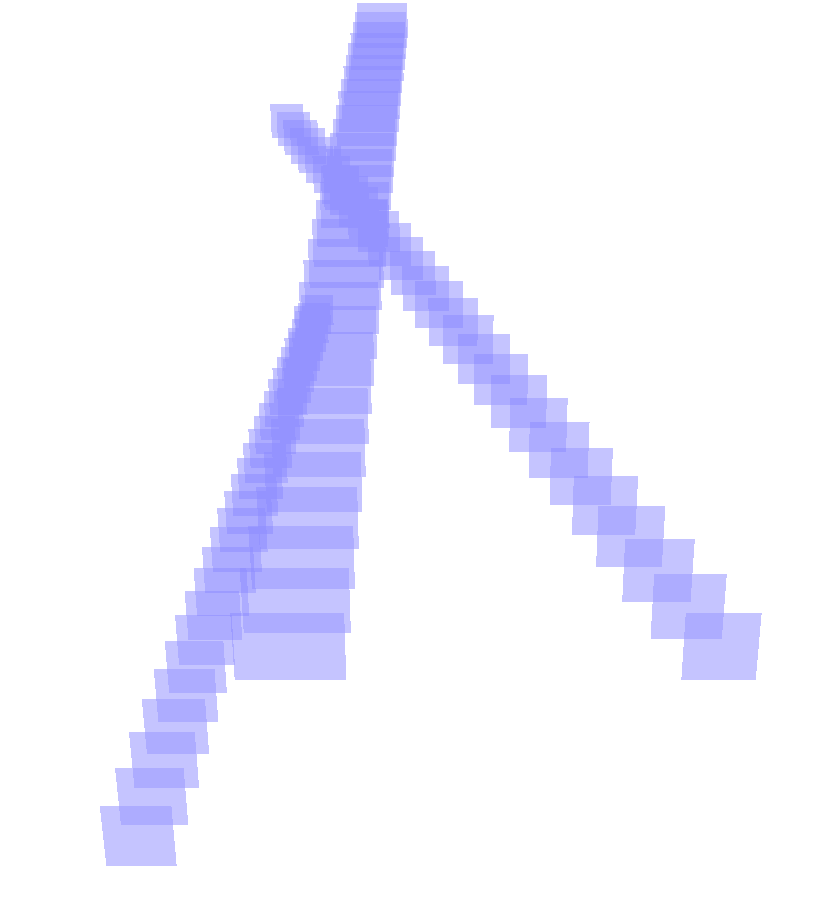}
\caption{3D graphical invariant representation}
\label{fig:3dinv}
\end{figure}

\paragraph{Visualization}
In our framework single invariants and scenarios consisting of multiple
invariants can be graphically depicted as demonstrated in the figures
throughout the report. In Figure~\ref{fig:3dinv}
the x and y coordinates of three objects' spaces are plotted
against the time. 

\section{Conclusion and Future Work}
\label{sec:concl}

We presented work towards a unified approach to
cyber-physical systems and component-based software development by
motivating a new framework 
concentrating on spatial aspects of cyber-physical systems. 
We described a process
  for checking properties of our models and described the approach
  using different examples.

Different directions for future work and ongoing work comprise: 
1) Taking the parallel checking of our models into
account and implementing it for a suitable platform.
2) Offering SOA based services for spatial modeling and checking.
3) Building a spatial behavioral type system for components following
our previous ideas \cite{blech1,blech2}.

Another more academic area of future work is reasoning about spatial properties in
proof assistants like Coq\footnote{\url{http://coq.inria.fr/}} or
Isabelle \cite{isabelle}. Here, we are aiming at 1) connecting proof
assistants as additional reasoning tools to manually verify concrete properties at an
invariant or verification condition level that are undecidable 2) derive meta-results on
the process of getting from models to invariants and verification conditions.
\bibliographystyle{plain}

\end{document}